\documentclass[runningheads]{llncs}
\usepackage{fullpage}
\usepackage{microtype}
\usepackage{amsmath,amssymb,amsfonts}
\usepackage{booktabs,tabularx}
\usepackage{url,xspace,soul,wrapfig}
\usepackage{graphicx,paralist,enumerate}
\usepackage{subfigure}
\usepackage[basic]{complexity}
\usepackage{hyperref}

\usepackage[nameinlink,capitalise]{cleveref}

\newcommand{\bigblacktriangle}{\protect\scalebox{1.2}{\ensuremath{\blacktriangle}}}

\Crefname{observation}{Observation}{Observations}
\Crefname{algorithm}{Algorithm}{Algorithms}
\Crefname{section}{Section}{Sections}
\Crefname{observation}{Observation}{Observations}
\Crefname{lemma}{Lemma}{Lemmas}
\Crefname{figure}{Fig.}{Figs.}
\Crefname{figure}{Fig.}{Figs.}
\Crefname{enumi}{Property}{Properties}
\Crefname{property}{Property}{Properties}

\DeclareMathOperator{\psn}{psn}
\newcommand{\T}{\mathcal T}

\usepackage{color}
\definecolor{realblue}{rgb}{0,0,1}
\definecolor{darkerblue}{rgb}{0.094,0.455,0.804}
\definecolor{darkblue}{rgb}{0.063,0.306,0.545}
\definecolor{red}{rgb}{1,0,0}
\definecolor{green}{rgb}{0,0.588,0.509}
\definecolor{orange}{rgb}{0.903,0.739,0.382}
\definecolor{realred}{rgb}{1,0,0}
\definecolor{purple}{rgb}{0.627,0.125,0.941}
\definecolor{pink}{rgb}{1,0.753,0.796}

\usepackage{ amssymb }

\newcommand{\orange}[1]{{{\textcolor{orange}{#1}\xspace}}}
\newcommand{\blue}[1]{{{\textcolor{blue}{#1}\xspace}}}
\newcommand{\red}[1]{{{\textcolor{red}{#1}\xspace}}}

\newcommand{\purple}[1]{{{\textcolor{purple}{#1}\xspace}}}

\newcommand{\leftx}{\ensuremath{\scalebox{.8}{$\nwarrow$}}}
\newcommand{\rightx}{\ensuremath{\scalebox{.8}{$\nearrow$}}}

\hypersetup{colorlinks,linkcolor={blue},citecolor={blue},urlcolor={blue}}
\renewcommand{\paragraph}[1]{\smallskip
\par\noindent {#1}~}
\renewcommand{\subsubsection}[1]{\smallskip
\par\noindent \textbf{#1}~}

\begin{document}

\title{Planar Drawings with Few Slopes\\ of Halin Graphs and Nested Pseudotrees
}

\titlerunning{Planar Drawings with Few Slopes of Halin Graphs and Nested Pseudotrees}

\authorrunning{Chaplick, Da Lozzo, Di Giacomo, Liotta, Montecchiani}

\author{Steven~Chaplick$^*$, Giordano~{Da~Lozzo}$^\diamond$, Emilio~{Di~Giacomo}$^\circ$, Giuseppe~Liotta$^\circ$, Fabrizio~Montecchiani$^\circ$}

\institute{
$^*$~Maastricht University, Maastricht, The Netherlands\\
\email{s.chaplick@maastrichtuniversity.nl}\\
$^\diamond$~Roma Tre University, Rome, Italy\\
\email{giordano.dalozzo@uniroma3.it}\\
$^\circ$~Dipartimento di Ingegneria, University of Perugia, Italy \\
\email{\{emilio.digiacomo,giuseppe.liotta,fabrizio.montecchiani\}@unipg.it}          
}

\maketitle

\begin{abstract}
The \emph{planar slope number} $\psn(G)$ of a planar graph $G$ is the minimum number of edge slopes in a planar straight-line drawing of $G$. It is known that $\psn(G) \in O(c^{\Delta})$ for every planar graph $G$ of maximum  degree $\Delta$. This upper bound has been improved to $O(\Delta^5)$ if $G$ has treewidth three, and to $O(\Delta)$ if $G$ has treewidth two. In this paper we prove $\psn(G) \leq \max\{4,\Delta\}$ when $G$ is a Halin graph, and thus has treewidth three. Furthermore, we present the first polynomial upper bound on the planar slope number for a family of graphs having treewidth four. Namely  we show that $O(\Delta^2)$ slopes suffice for nested pseudotrees. 
\end{abstract}

\section{Introduction}
Minimizing the number of slopes used by the edge segments of a straight-line graph drawing is a well-studied problem, which has received notable attention since its introduction by Wade and Chu~\cite{Wade01011994}. A break-through result by Keszegh, Pach and P{\'a}lv{\"o}lgyi~\cite{DBLP:journals/siamdm/KeszeghPP13} states that every planar graph of maximum degree $\Delta$ admits a planar straight-line drawing using at most $2^{O(\Delta)}$ slopes. That is, the \emph{planar slope number} of planar graphs is bounded by a function of $\Delta$, which answers a question of Dujmovi\'c et al.~\cite{DBLP:journals/comgeo/DujmovicESW07}. In contrast, the slope number of non-planar graphs has been shown to be unbounded (with respect to $\Delta$) even for $\Delta = 5$~\cite{DBLP:journals/combinatorics/BaratMW06,DBLP:journals/combinatorics/PachP06}. Besides the above mentioned upper bound,  Keszegh et al.~\cite{DBLP:journals/siamdm/KeszeghPP13} also prove a lower bound of $3\Delta - 6$, leaving as an open problem to reduce the large gap between upper and lower bounds on the planar slope number of~planar~graphs.

\begin{wrapfigure}[17]{R}{0.45\columnwidth}
	\vspace*{-.5cm}
	\centering
	\includegraphics[width=0.45\columnwidth,page=2]{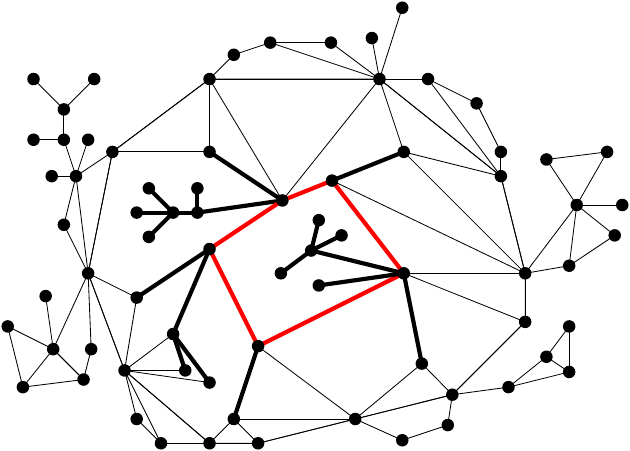}
	\caption{A nested pseudotree: the edges of its pseudotree are bold and the cycle of its pseudotree is red. %
	}
	\label{fig:nested-pseudo-example}
\end{wrapfigure}
The open problem by Keszegh et al. motivated a great research effort to establish improvements for subclasses of planar graphs. Jel\'{\i}nek et al.~\cite{DBLP:journals/gc/JelinekJKLTV13} study planar partial $3$-trees and show that their planar slope number is at most $O(\Delta^5)$. Di~Giacomo et al.~\cite{DBLP:journals/jgaa/GiacomoLM15} study a subclass of planar partial $3$-trees (those admitting an outer $1$-planar drawing) and present an $O(\Delta^2)$ upper bound for the planar slope number of these graphs.
Lenhart et al.~\cite{DBLP:conf/gd/LenhartLMN13} prove that the planar slope number of a partial $2$-tree is at most $2\Delta$ (and some partial $2$-trees require at least $\Delta$ slopes). Knauer et al.~\cite{DBLP:journals/comgeo/KnauerMW14} focus on outerplanar graphs (a subclass of partial $2$-trees) and establish a tight bound of $\Delta - 1$ for the (outer)planar slope number of this graph class. Di Giacomo et al.~\cite{DBLP:journals/tcs/GiacomoLM18} prove that the planar slope number of planar graphs of maximum degree three is four. Finally, the problem of computing planar drawings with few slopes has also been studied in the setting where the edges are polygonal chains rather than straight-line segments~\cite{DBLP:journals/algorithmica/AngeliniBLM19,DBLP:journals/algorithmica/new,DBLP:journals/comgeo/GiacomoLM20,DBLP:journals/siamdm/KeszeghPP13,DBLP:journals/jgaa/KindermannMSS21,DBLP:conf/latin/KnauerW16}.

An algorithmic strategy to tackle the study of the planar slope number problem can be based on a \emph{peeling-into-levels} approach. This approach has been successfully used to address the planar slope number problem for planar $3$-trees~\cite{DBLP:journals/gc/JelinekJKLTV13}, as well as to solve several other algorithmic problems on (near) planar graphs, including determining their pagenumber~\cite{DBLP:conf/compgeom/BekosLGGMR20,DBLP:journals/algorithmica/BekosBKR17,DBLP:journals/jgaa/DujmovicF18,DBLP:journals/jcss/Yannakakis89}, computing their girth~\cite{DBLP:journals/siamcomp/ChangL13}, and constructing radial drawings~\cite{DBLP:journals/jgaa/GiacomoDLM05}. In the peeling-into-levels approach the vertices of a plane graph (i.e., a planar graph with a given planar embedding) are partitioned into levels, based on their distance from the outer face. The vertices in each level induce an outerplane graph and two consecutive levels form a $2$-outerplane graph. One key ingredient is an algorithm that deals with a $2$-outerplane graph with possible constraints on one of the two levels. Another ingredient is an algorithm to extend a partial solution by introducing the vertices of new levels, while taking into account the constraints defined in the already-considered levels. Intrinsic in this approach is the construction of an embedding-preserving planar drawing of the input graph. The \emph{plane slope number} $\psn(G)$ of a plane graph $G$ is the minimum number of slopes used by the edge segments over all possible embedding-preserving planar straight-line drawings of $G$. Clearly, the planar slope number of $G$ is upper bounded by its plane slope number.

In an attempt to exploit the peeling-into-levels approach to prove a polynomial upper bound on the plane slope number of general plane graphs, one must be able to show a polynomial bound on the plane slope number of $2$-outerplane graphs. In this paper we take a first step in this direction by focusing on a meaningful subfamily of $2$-outerplane graphs, namely the nested pseudotrees. A \emph{nested pseudotree} is a graph with a planar embedding such that when removing the vertices of the outer face one is left with a pseudotree, that is, a connected graph with at most one cycle. See \Cref{fig:nested-pseudo-example} for an example. The family of nested pseudotrees generalizes the well studied $2$-outerplane simply nested graphs and properly includes the Halin graphs~\cite{halin-71}, the cycle-trees~\cite{DBLP:conf/isaac/LozzoDEJ17}, and the cycle-cycles~\cite{DBLP:conf/isaac/LozzoDEJ17}. Simply nested graphs were first introduced by Cimikowski~\cite{DBLP:journals/ipl/Cimikowski90}, who proved that the inner-triangulated ones are Hamiltonian, and have been extensively studied
in various contexts, such as universal point sets~\cite{DBLP:journals/dcg/AngeliniBBKMRS18,DBLP:conf/gd/AngeliniBKMRS11}, square-contact representations~\cite{DBLP:conf/isaac/LozzoDEJ17}, and clustered planarity~\cite{DBLP:conf/wg/LozzoEG018}.
Generally, nested pseudotrees have treewidth four and, as such, the best prior upper bound on their planar slope number is the one by Keszegh et al., which is exponential in $\Delta$. Halin graphs and cycle-trees have instead treewidth three, and therefore the previously known upper bound for these graphs is $O(\Delta^5)$, as shown by Jel\'{\i}nek et al.~\cite{DBLP:journals/gc/JelinekJKLTV13}. We prove significantly better upper bounds for all the above mentioned graph classes. Our main results are the following.

\begin{theorem}\label{thm:main}
Every nested pseudotree $G$ with maximum degree $\Delta$ has $\psn(G) \in O(\Delta^2)$.
\end{theorem}

\begin{theorem}\label{thm:main-halin}
Every Halin graph $G$ with maximum degree $\Delta$ different from $K_4$ has $\psn(G) \leq \max\{4,\Delta\}$.
\end{theorem}

\noindent The proofs of \cref{thm:main,thm:main-halin} are constructive. We first consider the case that $G$ is a cycle-tree and we show a recursive algorithm that computes a planar straight-line drawing of $G$  using $O(\Delta^2)$ slopes (\cref{se:cycle-tree}). The algorithm first considers $3$-connected instances which are treated by means of a suitable data structure called SPQ-tree~\cite{DBLP:conf/isaac/LozzoDEJ17}. The case of general nested pseudotree is then handled in \cref{se:nested} where the input graph $G$ is transformed into a cycle-tree $G'$ by removing an edge $e$; $G'$ is drawn with the algorithm of \cref{se:cycle-tree} and the invariants that we maintain in the construction are exploited to reintroduce $e$ in the computed drawing so to obtain a new drawing that still uses $O(\Delta^2)$ slopes. The technique for cycle-trees described in \cref{se:cycle-tree} gives an upper bound of $12 \Delta+10$ when applied to a Halin graph; in \Cref{se:halin} we prove for this family the finer bound stated in \Cref{thm:main-halin}. \cref{se:open} discusses some open problems.

\section{Preliminaries} \label{se:prel}

We assume familiarity with standard graph theoretic and graph drawing notions (see, e.g.,~\cite{DBLP:books/ph/BattistaETT99,DBLP:books/daglib/0030488}). Let $G$ be a graph and let $v$ be a vertex of $G$; let $\deg_G(v)$ denote the degree of vertex $v$ of $G$. The \emph{degree} $\Delta(G)$ of $G$ is $\max_{v \in G} \deg_G(v)$. When clear from the context, we omit the specification of $G$ in the above notation and say that $G$ is a \emph{degree-$\Delta$} graph. 

\subsubsection{Drawings and Embeddings.} A \emph{drawing} $\Gamma$ of a graph $G$ is a mapping of the vertices of $G$ to distinct points of the plane, and of the edges of $G$ to Jordan arcs connecting their corresponding endpoints but not passing through any other vertex. In the remainder of the paper, if it leads to no confusion, in notation and terminology we make no distinction between a vertex of $G$ and the corresponding point of $\Gamma$ and between an edge of $G$ and the corresponding arc of $\Gamma$. Drawing~$\Gamma$ is \emph{straight-line} if its edges are straight-line segments. A drawing is \emph{planar} if no two edges intersect, except at a common endpoint, if any. A \emph{planar graph} is a graph that admits a planar drawing. A planar drawing subdivides the plane into topologically connected regions, called \emph{faces}. The infinite region is called the \emph{outer face}; any other face is an \emph{inner face}. A \emph{planar embedding} of a planar graph is an equivalence class of topologically equivalent (i.e., isotopic) planar drawings of $G$. A planar embedding of a connected planar graph can be described by the clockwise circular order of the edges around each vertex together with the choice of the outer face. A planar graph with a given planar embedding is a \emph{plane graph}. A \emph{plane drawing} of a plane graph $G$ is a planar drawing of $G$ that preserves the planar embedding of $G$.

The \emph{slope} of a line $\ell$ is the smallest angle $\alpha \in [0,\pi)$ such that $\ell$ can be made horizontal by a clockwise rotation by $\alpha$. The \emph{slope} of a segment is the slope of the line containing it. 
Let $G$ be a plane graph and let $\Gamma$ be a plane straight-line drawing of $G$. The \emph{plane slope number} $\psn(\Gamma)$ of $\Gamma$ is the number of distinct slopes used by the edges of $G$ in $\Gamma$. The \emph{plane slope number} $\psn(G)$ of $G$ is the minimum~$\psn(\Gamma)$ over all planar straight-line drawings $\Gamma$ of $G$. If $G$ has degree $\Delta$, then clearly $\psn(G) \geq \lceil \Delta/2 \rceil$, as in any straight-line drawing the same slope can be used by at most two edges incident to the same vertex.

\begin{figure}[htb]
	\centering
	\subfigure[]{\label{fig:2-trees-construction-a}\includegraphics[width=0.18\columnwidth,page=1]{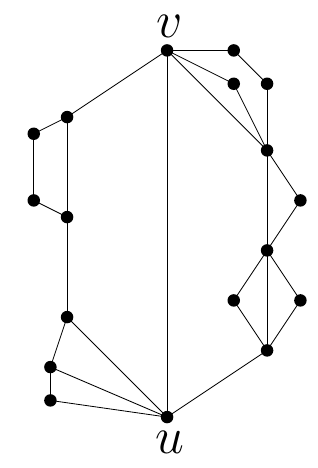}}
	\subfigure[]{\label{fig:2-trees-construction-b}\includegraphics[width=0.35\columnwidth,page=2]{figs/2-trees-construction}}
	\subfigure[]{\label{fig:2-trees-construction-c}\includegraphics[width=0.35\columnwidth,page=3]{figs/2-trees-construction}}
	\caption{(a) A plane partial $2$-tree with a distinguished edge $(u,v)$;  (b) A rhombus $\blacklozenge(a b c d)$ with $s=0$ and angles at $a$ and $c$ equal to $\beta$; (c) a plane straight-line drawing of $G$ inside $\blacklozenge(a b c d)$; the slope set $\mathcal L(\beta,s,\Delta)$ used is shown in the figure.}
	\label{fig:2-trees-construction}
\end{figure}

The following theorem rephrases a result proved in~\cite{DBLP:conf/gd/LenhartLMN13}. \Cref{fig:2-trees-construction} shows an example of the construction.

\begin{theorem}[\cite{DBLP:conf/gd/LenhartLMN13}]\label{prop:series-parallel}
	Let $G$ be a degree-$\Delta$ plane partial $2$-tree and let $(u,v)$ be a distinguished edge of $G$.
	Let $0 < \beta < \frac{\pi}{2}$, let $s$ be a given slope, and let $\blacklozenge(a b c d)$ be any rhombus whose longer diagonal $\overline{ac}$ has slope $s$ and such that the interior angles at $a$ and $c$ are equal to $\beta$. 
	There exists a set $\mathcal L(\beta,s,\Delta)$ of $O(\Delta)$ slopes such that $G$ admits a plane straight-line drawing inside $\blacklozenge(a b c d)$ using the slopes in $\mathcal L(\beta,s,\Delta)$ and such that~$a \equiv v$ and $c\equiv u$.
\end{theorem}

\subsubsection{Nested pseudotrees.}
\begin{figure}[htb]
	\centering
	\subfigure[]{\label{fig:families-a}\includegraphics[width=0.42\columnwidth,page=1]{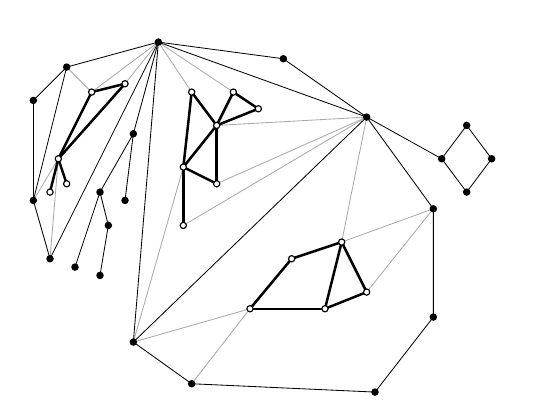}}
    \subfigure[]{\label{fig:families-b}\includegraphics[width=0.42\columnwidth,page=2]{figs/families}}
    \subfigure[]{\label{fig:families-c}\includegraphics[width=0.42\columnwidth,page=3]{figs/families}}
    \subfigure[]{\label{fig:families-d}\includegraphics[width=0.42\columnwidth,page=4]{figs/families}} 
\caption{Examples of $2$-outerplane graphs; In all figures, the external vertices are black and the internal ones are white. The edges connecting external vertices are thin black; the edges connecting internal vertices are bold; the edges connecting an internal and an external vertex are grey. 
 (b), (c), and (d) are simply nested $2$-outerplane graphs; (b) is a cycle-cycle; (c) and (d) are Cycle-trees; (d) is a Halin graph.}
	\label{fig:families}
\end{figure}

A planar drawing of a graph is \emph{outerplanar} if all the vertices are incident to the outer face,
and \emph{$2$-outerplanar} if removing the vertices of the outer face yields an outerplanar graph.
A graph is \emph{$2$-outerplanar} (\emph{outerplanar}) if it admits a $2$-outerplanar drawing (outerplanar drawing). See \Cref{fig:families-a,fig:families-b,fig:families-c,fig:families-d} for examples of $2$-outerplanar graphs. In a $2$-outerplanar drawing, vertices incident to the outer face are called \emph{external}, and all other vertices are \emph{internal}.
A \emph{$2$-outerplane} graph is a $2$-outerplanar graph with a planar embedding inherited from a $2$-outerplanar drawing.
A $2$-outerplane graph is \emph{simply nested} if its external vertices induce a chordless cycle and its internal vertices induce either a chordless cycle or a tree. See, for example, \Cref{fig:families-b,fig:families-c,fig:families-d}
As in~\cite{DBLP:conf/isaac/LozzoDEJ17}, we refer to a simply nested graph whose internal vertices induce a chordless cycle or a tree as a \emph{cycle-cycle}  or a \emph{cycle-tree}, respectively. See \Cref{fig:families-b} for an example of a cycle-cycle and \Cref{fig:families-c,fig:families-d} for two examples of cycle-trees. 
A \emph{Halin graph} is a $3$-connected plane graph $G$ such that, by removing the edges incident to the outer face, one gets a tree whose internal vertices have degree at least $3$ and whose leaves are incident to the outerface of $G$. See \Cref{fig:families-d} for an example of a Halin graph. By definition, Halin graphs are a subfamily of the cycle-trees.
A \emph{pseudotree} is a connected graph containing at most one cycle.
A \emph{nested pseudotree} is a topological graph such that removing the vertices on the outer face yields a non-empty pseudotree. See~\Cref{fig:nested-pseudo-example} for an example of a nested pseudotree.  Note that the external vertices of a nested pseudotree need not induce a chordless cycle. In fact, the outer boundary is a closed walk. By definition, nested pseudotrees generalize $2$-outerplane simply nested graphs. Moreover, this class includes some graphs of treewidth $4$, as formalized in the following.

\begin{theorem}\label{thm:treewidth}
Nested pseudotrees have treewidth at most $4$, which is tight.
\end{theorem}
\begin{proof} 
	\smallskip\noindent{\em Lower bound.} The graph of the octahedron is a nested pseudotree whose cycle and pseudotree are both triangles. This graph is one of the forbidden minors for treewidth-$3$ graphs~\cite{DBLP:journals/dm/ArnborgPC90}.	Hence, there is a nested pseudotree with treewidth~at~least~$4$.

	\smallskip\noindent{\em Upper bound.}  	First, we show that each cycle-tree has treewidth at most~$3$, and then we  improve this bound to show that each nested pseudotree has treewidth~at~most~$4$. 

	Since each cycle-tree has \emph{radius} $r=1$ (defined as the maximum distance of an inner face from the outer face), it follows from a theorem of Robertson and Seymour~\cite{DBLP:journals/jct/RobertsonS84} that they have treewidth at most~$3r+1=4$. However, we can prove, more strongly, that any cycle-tree $G$ has treewidth at most~$3$. To this aim we prove that $G$ has a tree decomposition of width~$3$. Let $C$ be the cycle induced by the external vertices of $G$ and let $T$ be the tree induced by the internal vertices of $G$. We can assume that $G$ is $2$-connected. If $G$ is not $2$-connected, then it has a $2$-connected component $B$ that contains all the vertices of $C$; any other $2$-connected component only contains vertices of $T$ and hence it is an edge. Since the treewidth of a graph is the maximum treewidth of its $2$-connected components~\cite{DBLP:journals/tcs/Bodlaender98}, we can concentrate on the component $B$. The proof is by induction on the number of edges in $T$. If $T$ has no edge, then it has only one vertex and $G$ is a subgraph of a wheel graph. Since a wheel graph is a Halin graph, it has treewidth~$3$~\cite{DBLP:journals/tcs/Bodlaender98}.
	
	\begin{figure}[h!]
		\centering
		\includegraphics[width=.7\textwidth]{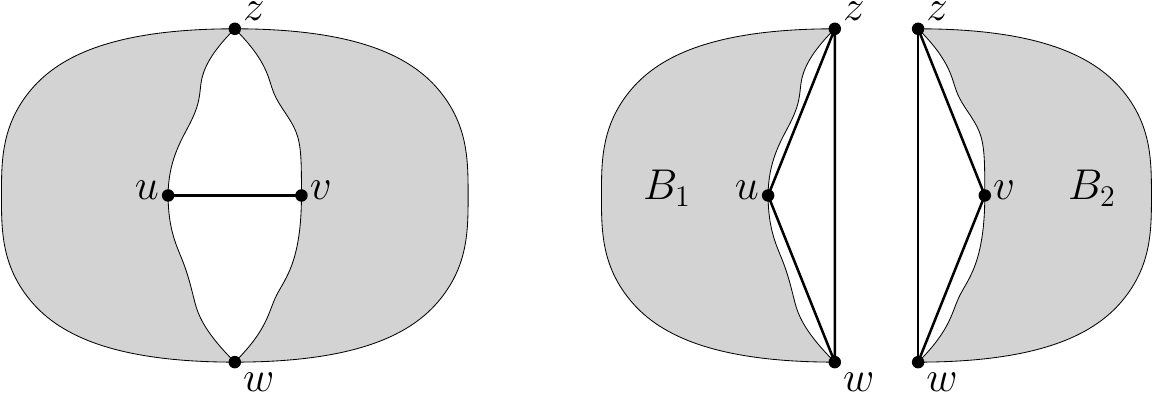}
		\caption{Illustration for the proof of \Cref{thm:treewidth}}
		\label{fig:cycle-trees}
	\end{figure}
	
	Suppose now that $T$ has at least one edge $(u,v)$. This edge is shared by two faces of $G$, each of which contains at least one external vertex. Let $w$ and $z$ be two external vertices, one for each of the two faces. Removing $(u,v)$, the pair $\{w,z\}$ becomes a $2$-cut (see \Cref{fig:cycle-trees}), which splits the graphs into two components, one having $w$, $z$, and $u$ on the outer boundary, call it $B_1$, and one having $w$, $z$, and~$v$ on the outer boundary, call it $B_2$. We add to $B_1$ the edges $(w,z)$, $(z,u)$, and $(u,w)$, if they do not exist; analogously, we add to $B_2$ the edges $(w,z)$, $(z,v)$, and $(v,w)$, if they do not exist. After this addition both $B_1$ and $B_2$ are cycle-trees whose trees have at least one edge less that the tree of $G$. By induction they admit a tree decomposition of width $3$. Since $B_1$ contains the $3$-cycle $(w,z)$, $(z,u)$, and~$(u,w)$, in the tree decomposition of $B_1$ there is a bag $X_1$ containing the three vertices $u$, $w$ and $z$. Analogously, in the tree decomposition of $B_2$ there is a bag $X_2$ containing the three vertices $v$, $w$, and $z$. We combine these two tree decompositions by adding a new bag containing the vertices $u$, $v$, $w$, and $z$ and connecting it to both $X_1$ and $X_2$. This results in a tree decomposition of $G$, with width $3$.
	
	We will use this result to establish that the treewidth of any nested pseudotree is at most $4$. First, notice that a cycle-psuedotree is simply a cycle-tree plus one edge. 
	Moreover, adding one edge to any graph increases the treewidth by at most one. 
	Thus, since we have shown that every cycle-tree has treewidth at most 3, each cycle-peseudotree has treewidth at most 4. 
	
	Finally, we extend this bound to each nested pseudotree. 
	Consider any nested pseudotree $G$. 
	Observe that $G$ consists of a cycle-pseudotree $H$ together with a (possibly empty) set of partial $2$-trees hanging from $2$-cuts formed by edges of the chordless cycle containing the psuedotree. 
	Consequently, any tree decomposition of $H$ with width $t$ can be extended to a tree decomposition of $G$ where the width is $\max\{2,t\}$. 
	Thus, since $H$ has treewidth at most~$4$, we have that $G$ also has treewidth at most~$4$. 
\qed\end{proof}

\section[Cycle-Trees and Proof of Theorem 1.1]{Cycle-Trees and Proof of \Cref{thm:main-halin}}\label{se:cycle-tree}

In this section, we consider cycle-trees and prove that their plane slope number is~$O(\Delta^2)$ in general and $O(\Delta)$ for Halin graphs. A degree-$2$ vertex $v$ of a cycle-tree~$G$ whose neighbors are $x$ and $y$ is \emph{contractible} if $(x,y)$ is not an edge of $G$, and if deleting $v$ and adding the edge $(x,y)$ yields a cycle-tree; this operations is the \emph{contraction} of $v$.
A cycle-tree $G$ is \emph{irreducible} if it contains no contractible vertex.

\begin{lemma}\label{le:nice}
For every degree-$\Delta$ cycle-tree $G$ and irreducible cycle-tree $G'$ obtained from $G$ by any sequence of contractions, $\psn(G) \leq \psn(G')$.
\end{lemma}
\begin{proof}
	First, $G'$ has at most degree~$\Delta$, as each contraction does not increase the degree of any vertex. Let $\Gamma'$ be a plane straight-line drawing of $G'$. A plane straight-line drawing $\Gamma$ of $G$ can be obtained from $\Gamma'$ by subdividing the edges that stemmed from the contraction operations. Clearly, $\psn(\Gamma)=\psn(\Gamma')$, and consequently~$\psn(G) \leq \psn(G')$.
\qed\end{proof}

By \cref{le:nice}, without loss of generality, the considered cycle-trees will have no contractible vertices. 
Furthermore, if the outer face of an irreducible $2$-connected cycle-tree $G$ of degree $\Delta$ has size $k \geq 3$, then the number of edges of $G$ is $O(k \,\Delta)$, which implies that $\psn(G) \in O(\Delta)$ if $k$ is constant. This observation allows us to assume $k>3$ for $2$-connected instances, which will simplify the description.

\subsection{$3$-Connected Instances}\label{ss:3connected}
\begin{figure}[htb]
	\centering
	\subfigure[]{\label{fig:path-tree-a}\includegraphics[scale=0.6,page=1]{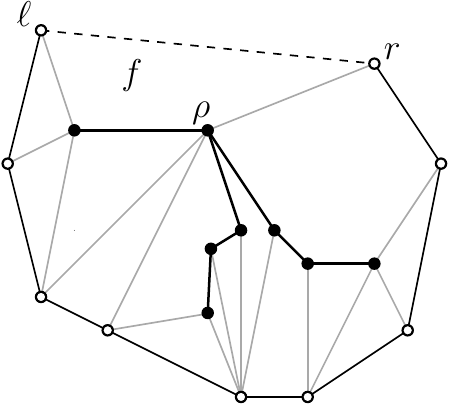}}
	\hfil
	\subfigure[]{\label{fig:path-tree-b}\includegraphics[scale=0.6,page=2]{figs/path-tree}}
	\caption{(a) An example of a path-tree $G$ (solid edges). Vertex $\ell$ is the leftmost path vertex and $r$ is the rightmost path vertex of $G$. The path of $G$ is shown with white vertices and black solid edges, while the tree of $G$ is shown with black vertices and bold edges. (b) The path-tree~$G$ rooted at $\rho$.}
	\label{fig:path-tree}
\end{figure}

		A \emph{path-tree} is a plane graph $G$ that can be augmented to a cycle-tree~$G'$ by adding an edge $e=(\ell,r)$ to its outer face. \Cref{fig:path-tree-a} shows an example of a path-tree where the edge $(\ell,r)$ is the dashed edge. Suppose that, in a clockwise walk along the outer face of $G'$, edge $e$ is traversed from $\ell$ to $r$; then $\ell$ is the \emph{leftmost path-vertex} and $v$ is the \emph{rightmost path-vertex} of $G$. All vertices in the outer face of $G'$ are \emph{path-vertices}, while the other vertices are \emph{tree-vertices}.  The path induced by the path-vertices of $G$ is the \emph{path of $G$}. Analogously, the tree induced by the tree-vertices of $G$ is the \emph{tree of $G$}. In \Cref{fig:path-tree} the path of $G$ is shown with white vertices and black solid edges, while the tree of $G$ is shown with black vertices and bold edges. Let $f$ be the internal face of $G'$ that contains edge $e$.  The path-tree $G$ can be \emph{rooted} at any tree-vertex $\rho$ on the boundary of $f$; then vertex $\rho$ becomes the \emph{root} of $G$. \Cref{fig:path-tree-b} shows the path-tree of \Cref{fig:path-tree-a} rooted at a vertex $\rho$. If~$G$ is rooted at $\rho$, then the tree of $G$ is also rooted at $\rho$.   
A rooted path-tree with root~$\rho$, leftmost path-vertex $\ell$, and rightmost path-vertex $r$ is \emph{almost-$3$-connected} if it becomes $3$-connected by adding the edges $(\rho,\ell)$, $(\rho, r)$, and $(\ell, r)$, if missing. For example, the path-tree of \Cref{fig:path-tree} is almost-3-connected.

\begin{figure}[htb!]
	\centering
		\includegraphics[scale=0.7,page=2]{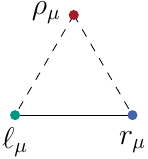}%
	\hfil
		\includegraphics[scale=0.7,page=3]{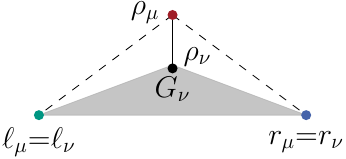}%
	\hfil
		\includegraphics[scale=0.7,page=4]{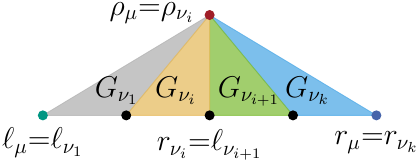}%
	\caption{Path-trees associated with a Q-node (left), an S-node (middle), and a P-node (right). Dashed edges may or may not exist. Shaded triangles represent smaller path-trees $G_{\nu_i}$ rooted at $\rho_i$, with leftmost path-vertex $\ell_{\nu_i}$ and rightmost path-vertex $r_{\nu_i}$.}
	\label{fig:nodes}
\end{figure}

\subsubsection{SPQ-decomposition of path-trees.} Let $G$ be an almost-$3$-connected path-tree rooted at $\rho$, with leftmost path-vertex $\ell$ and rightmost path-vertex $r$.
The \emph{SPQ-decomposition}  of $G$~\cite{DBLP:conf/isaac/LozzoDEJ17} constructs a tree $\mathcal{T}$, called the \emph{SPQ-tree} of $G$, whose nodes are of three different kinds: {\em S-}, {\em P-}, and {\em Q-nodes}.
Each node $\mu$ of $\mathcal{T}$ is associated with an almost-$3$-connected rooted path-tree $G_\mu$, called the \emph{pertinent graph} of~$\mu$.
To avoid special cases, we extend the definition of path-trees so to include graphs whose path is a single edge $(\ell,r)$ and whose tree consists of a single vertex $\rho$, possibly not adjacent to $\ell$ or $r$. As a consequence, we also extend the definition of almost-$3$-connected path-trees to graphs such that adding $(\rho, r)$, and $(\ell, r)$, if missing, yields a $3$-cycle.

\noindent\fbox{\sc Q-node:} the pertinent graph $G_\mu$ of a \emph{Q-node} $\mu$ is an almost-$3$-connected rooted path-tree consisting of three vertices: one tree-vertex $\rho_\mu$ and two path-vertices~$\ell_\mu$ and~$r_\mu$.
Vertices $\rho_\mu$, $\ell_\mu$, and $r_\mu$ are the root, the leftmost path-vertex, and the rightmost path-vertex of $G_\mu$, respectively.
$G_\mu$ always has edge $(\ell_\mu, r_\mu)$, while $(\rho_\mu,\ell_\mu)$ and $(\rho_\mu, r_\mu)$ may not exist; see~\cref{fig:nodes}\blue{(left)}.

\noindent\fbox{\sc S-node:} the pertinent graph $G_\mu$ of an \emph{S-node} $\mu$ is an almost-$3$-connected rooted path-tree consisting of a root $\rho_\mu$  adjacent to the root $\rho_\nu$ of one almost-$3$-connected rooted path-tree $G_\nu$, and possibly to the leftmost path-vertex $\ell_\nu$ and to the rightmost path-vertex $r_\nu$ of $G_\nu$.
The node $\nu$ whose pertinent graph is $G_\nu$ is the unique child of $\mu$ in $\mathcal{T}$.
The leftmost and the rightmost path-vertices of $G_\mu$ are $\ell_\nu$ and $r_\nu$, respectively; see~\cref{fig:nodes}\blue{(middle)}.

\noindent\fbox{\sc P-node:} the pertinent graph $G_\mu$ of a \emph{P-node} $\mu$ is an almost-$3$-connected rooted path-tree obtained from almost-$3$-connected rooted path-trees $G_{\nu_1},\dots,G_{\nu_k}$, with $k>1$, as follows. First, the roots of $G_{\nu_1},\dots,G_{\nu_k}$ are identified into the root $\rho_\mu$ of~$G_\mu$. Second, the leftmost path-vertex of $G_{\nu_i}$ is identified with the rightmost path-vertex of $G_{\nu_{i-1}}$, for $i=2,\dots,k$.
The nodes $\nu_1,\dots,\nu_k$ whose pertinent graphs are~$G_{\nu_1},\dots,G_{\nu_k}$, respectively, are the children of $\mu$ in $\mathcal{T}$, and the left-to-right order in which they appear in $\cal T$ is $\nu_1,\dots,\nu_k$.
The leftmost and the rightmost path-vertices of $G_\mu$ are $\ell_{\nu_1}$ and $r_{\nu_k}$, respectively; see~\cref{fig:nodes}\blue{(right)}. 

The SPQ-tree $\cal T$ of $G$ is such that:
\begin{inparaenum}[\bf (i)]
	\item Q-nodes are leaves of $\mathcal{T}$.
	\item If the pertinent graph of an S-node $\mu$ contains neither $(\rho_\mu,\ell_\mu)$ nor $(\rho_\mu,r_\mu)$, then the parent of $\mu$ is a P-node.
	\item Every P-node has at most $2\Delta+1$ children.
\end{inparaenum}
\begin{figure}[t!]
\centering
\includegraphics[width=\textwidth,page=27]{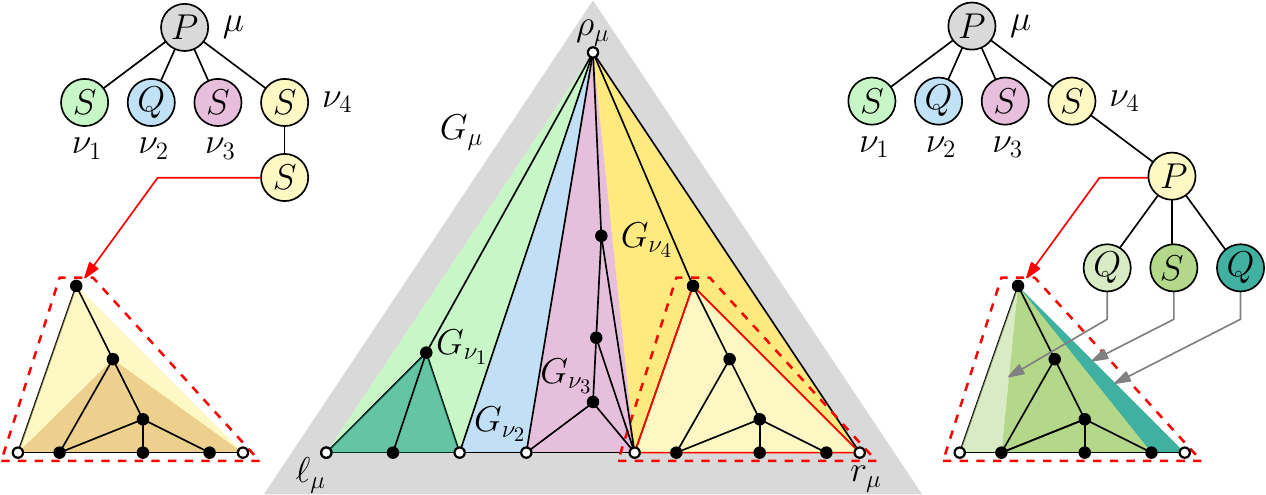}
\caption{Two alternative partial SPQ-trees of the almost-$3$-connected path-tree in the center: The child of node $\nu_4$ is an S-node on the left and a P-node~on~the~right.}
\label{fig:different-spq-trees}
\end{figure}
\cref{fig:different-spq-trees} provides two alternative SPQ-trees of the same graph.

For simplicity, we assume that the pertinent graphs of the children of a P-node~$\mu$ are induced subgraphs of $G_\mu$. This implies that if $G_\mu$ contains an edge~$(\rho_\mu,v)$, where $v$ is a path-vertex, then such an edge belongs to every child of $\mu$ whose pertinent graph contains $v$.

Let $\mu$ be a node of $\cal T$. The \emph{left path} of $\mu$ is the path directed from $\ell_\mu$ to $\rho_\mu$, consisting of edges belonging to the outer face of $G_\mu$, and not containing $r_\mu$. The definition of the  \emph{right path} of $\mu$ is symmetric.
Observe that, if $\mu$ is a Q-node whose pertinent graph $G_\mu$ does not contain the edge $(\rho_\mu,\ell_\mu)$, then the left path of $\mu$ is the empty path. Similarly, the right path of $\mu$ is the empty path if $G_\mu$ does not contain the edge $(\rho_\mu,r_\mu)$.

\noindent We say that an SPQ-tree is \emph{canonical} if each child of every P-node is either an S- or a Q-node.

\begin{lemma}\label{thm:path-tree-decomp}
	Every $n$-vertex almost-$3$-connected rooted path-tree $G$ admits a canonical SPQ-tree. Furthermore, a canonical SPQ-tree  of $G$ can be  computed in $O(n)$ time.
\end{lemma}

\begin{proof}
	By \cite{DBLP:conf/isaac/LozzoDEJ17}, any almost-$3$-connected rooted path-tree $G$ admits an SPQ-tree~$\cal T$. If $\cal T$ is not canonical,  it can be turned into a canonical SPQ-tree of $G$ as follows.  Let $\mu$ be a P-node in $\cal T$ with children $\nu_1,\dots,\nu_k$ such that $\nu_j$ is a P-node, with~$1 \leq j \leq k$. We remove $\nu_j$ from $\cal T$, and  we update the children of $\mu$ in $\cal T$ to be~$\nu_1,\dots,\nu_{j-1},\lambda_1,\dots,\lambda_h,\nu_{j+1},\dots,\nu_k$, where $\lambda_1,\dots,\lambda_h$ are the children of $\nu_j$ in~$\cal T$. We repeat this procedure until  $\cal T$ becomes canonical. In the following, we show how to directly compute a canonical SPQ-tree $\cal T$ of $G$ in linear time.

	Since $G$ is an almost-$3$-connected rooted path-tree, each internal face of $G$ is incident to exactly one or exactly two path vertices. In linear time, we label each internal face $f$ of $G$ with a list $L(f)$ containing 
	either the single path vertex or the two path vertices $f$ is incident to (in the left-to-right order in which they appear along the path of $G$). Furthermore we orient the edges incident to tree-vertices as follows: the tree edges are oriented from parent to children, while the edges connecting a tree-vertex to a path-vertex are oriented from the tree-vertex to the path-vertex.  Let $\rho$, $\ell$ and $r$ be the root, the left-most path-vertex, and the right-most path vertex of $G$. We compute $\cal T$ recursively. (\textbf{Base case}) If $G$ has exactly three vertices, namely $\rho$, $\ell$ and $r$, then $\cal T$ consists of a $Q$-node. Otherwise (\textbf{Recursive case}),  let $e_1, e_2, \dots, e_k$ be the outgoing edges of $\rho$ in left-to-right order and denote by $v_i$ the end-vertex of $e_i$ different from $\rho$. Let $f_1,\dots,f_{k+1}$ be the faces of $G$ incident to $\rho$ where $f_i$ is the face to the left of $e_i$ (for $i=1,2,\dots,k$) and $f_{k+1}$ is the face to the right of $e_k$.  Notice that $f_1$ and $f_{k+1}$ may coincide. Also, let $L_\rho$ be the list obtained by concatenating $\ell$, $v_1$, $L(f_2)$, $v_2$, $L(f_3)$, $\dots$,  $L(f_k)$, $v_k$, and~$r$ (i.e., we initialize $L_\rho = \ell \circ v_1 \circ L(f_2) \circ v_2 \circ L(f_3) \circ v_3 \circ \dots v_ {k-1}\circ L(f_k) \circ v_k \circ r$), and by suppressing repeated vertices. 
	Note that, $L_\rho$ contains all the path-vertices that are visible from $\rho$ along the path between $\ell$ and $r$. Let $w_1, w_2, \dots, w_h$ be such path-vertices. By construction, for each $i=1,\dots,h-1$, either there is exactly one tree-vertex between $w_i$ and $w_{i+1}$ in $L_{\rho}$, or they are consecutive.
	
	Suppose that $L_\rho$ contains exactly two path-vertices, namely $\ell$ and $r$. In this case the root of the SPQ-tree $\cal T$ of $G$ will be an $S$-node. Since $G$ has more than three vertices, there must be a tree-vertex $z$ between $\ell$ and $r$ in $L_{\rho}$. We recursively construct the SPQ-tree $\cal T'$ of $G'=G - \rho$, which is an almost-$3$-connected rooted path-tree rooted at $z$ with leftmost path-vertex $\ell$ and rightmost path-vertex $r$. The root of $\cal T$ is the S-node whose single child is the root of $\cal T'$.
	
	Suppose now that $L_\rho$ contains at least three path-vertices. In this case the root of the SPQ-tree $\cal T$ of $G$ will be a $P$-node. For every pair $w_i$, $w_{i+1}$, we recursively construct the SPQ-tree $\mathcal T_i$ of an almost-$3$-connected rooted path-tree $G_i$ rooted at~$\rho$ with leftmost path-vertex $w_i$ and rightmost path-vertex~$w_{i+1}$. If there is no tree-vertex between $w_i$ and $w_{i+1}$, then $G_i$ is the subgraph of $G$ induced by $\rho$, $w_i$ and~$w_{i+1}$. If there is a tree-vertex $z_i$ between $w_i$ and $w_{i+1}$, then $G_i$ is the subgraph of $G$ induced by $\rho$, $w_i$, $w_{i+1}$, the path-vertices of $G$ between $w_i$ and $w_{i+1}$, and all the tree-vertices that are descendants of $z_i$ (including $z_i$). Observe that each $G_i$ is an almost-$3$-connected rooted path-tree and $\bigcup_{i=1}^{h-1}G_i$ coincides with $G$. Furthermore, in the first case above the root of $\mathcal T_i$ is a $Q$-node, while in the second case the root of $\mathcal T_i$ is an $S$-node. The root of $\cal T$ is the P-node whose children are the roots of the trees $\mathcal T_i$ (none of which are $P$-nodes), for $i=1,2,\dots,h-1$.  
	
	Concerning the running time, observe that when we construct the SPQ-tree~$\cal T$ of an almost-$3$-connected rooted path-tree rooted at $\rho$, the running time of the non-recursive part of the algorithm is $O(\deg(\rho))$ or $O(1)$ if the root of $\cal T$ is a P-node or a S-node, respectively. Furthermore, each tree-vertex $\rho$ can occur at most once as the root of the pertinent graph of a P-node and $O(\deg(\rho))$ many times as the root of the pertinent graph of an S- or Q-node. It follows that the overall running time is $O(n)$.\qed\end{proof}

Based on \Cref{thm:path-tree-decomp}, in the remainder we shall assume that our SPQ-trees are canonical.
The cornerstone of our contribution is a construction for almost-$3$-connected rooted path-trees using $O(\Delta^2)$ slopes. We start by defining the slope set. Let $a$, $b$, and $c$ be points in $\mathbb{R}^2$; $\overline{a b}$ denotes the straight-line segment whose~endpoints are $a$ and $b$, and  $\bigblacktriangle(a b c)$ denotes the triangle whose corners are $a$, $b$, and~$c$.

\begin{figure}[b!]
	\centering
	\subfigure[\label{fig:main-slope-sets}]{\includegraphics[scale=.48,page=7]{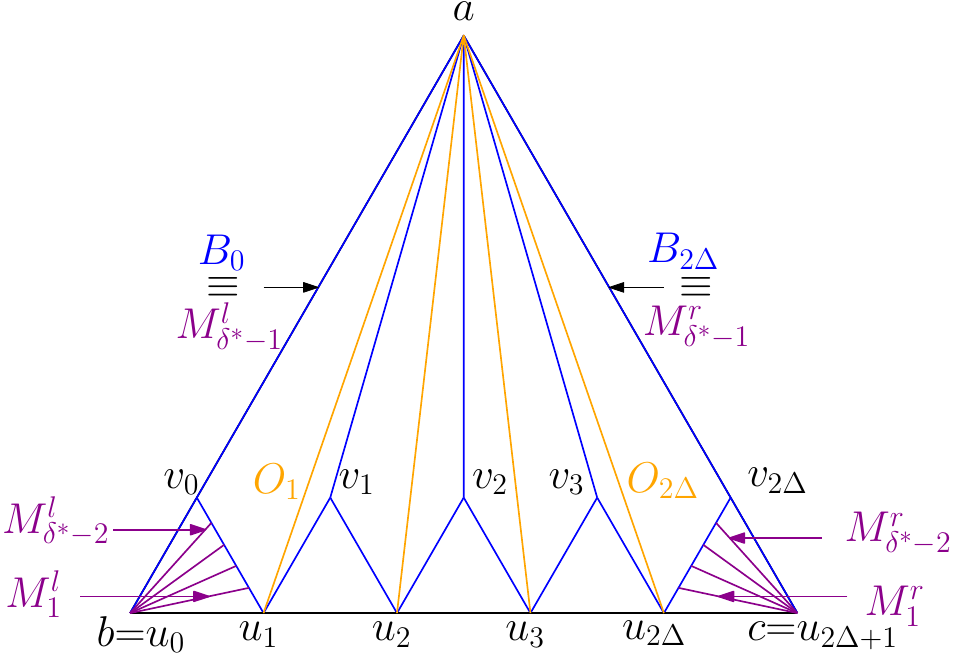}}
	\hfil
	\subfigure[\label{fig:main-central}]{\includegraphics[scale=.5,page=9]{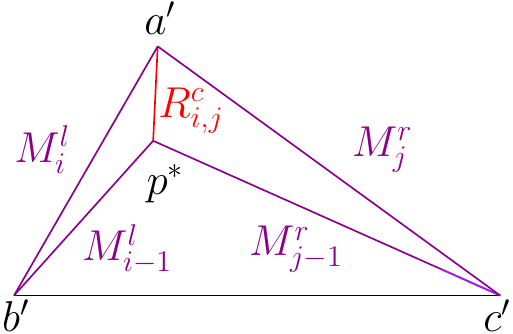}}
	\subfigure[]{
		\includegraphics[scale=.45,page=17]{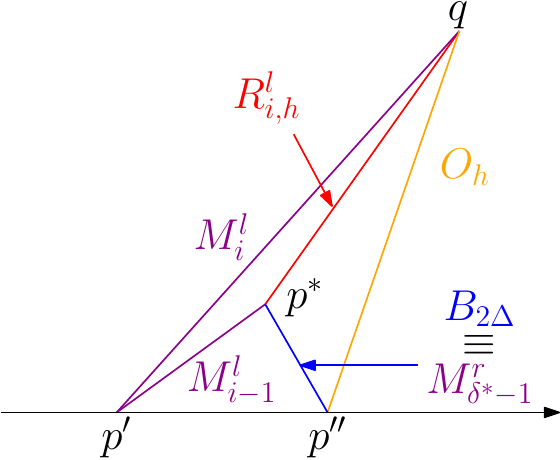}
		\label{fig:main-left-left}
	}
	\hfil
	\subfigure[]{
		\includegraphics[scale=.45,page=18]{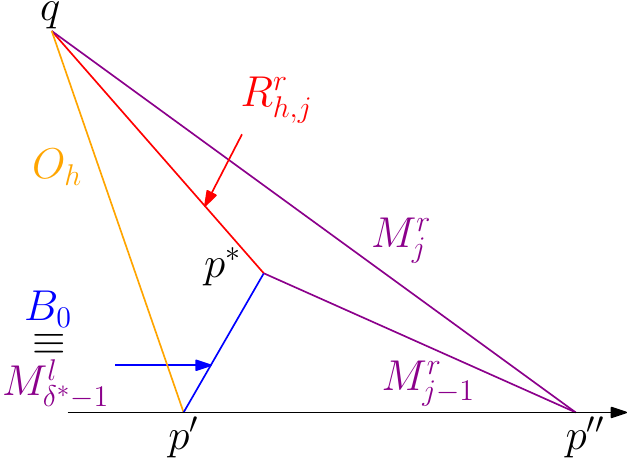}
		\label{fig:main-left-right}
	}
	\caption{(a) Black, orange, blue, left- and right-magenta slopes; (b) $c$-red~slope~$R^c_{i,j}$; (c) $l$-red slope $R^l_{i,h}$; and (d) $r$-red slope $R^r_{h,j}$.}
\end{figure}

\smallskip\noindent{\bf Slope set.}
Let $G$ be an almost-$3$-connected path-tree  and let $\cal T$ be an SPQ-tree of~$G$. For any node $\mu$ of $\cal T$ and for any path-vertex $v$ in $G_{\mu}$ we let $\delta_\mu(v)=\deg_{G_\mu}(v)$ and we let $\delta^*$ be the maximum $\delta_\mu(v)$ over all nodes $\mu$ and path-vertices $v$.
Consider the equilateral triangle $\bigblacktriangle(a b c)$ with vertices $a$, $b$, and $c$ in counter-clockwise order; refer to \cref{fig:main-slope-sets}. Assume that the side $\overline{b c}$ is horizontal, and that $a$ lies above $\overline{b c}$. Let $b=u_0,u_1,\dots,u_{2\Delta+1}=c$ be the $2\Delta+2$ equispaced points along $\overline{b c}$. We define the following slope sets:

\smallskip
\noindent{\em Black slope:} The slope $0$, i.e., the slope of an horizontal line.
\smallskip

\noindent{\em Orange slopes:} The \emph{$i$-th orange slope} $O_i$ is the slope of $\overline{a u_i}$, with $1 \leq i \leq 2\Delta$.
\smallskip

\noindent{\em Blue slopes:} The \emph{$i$-th blue slope} $B_i$ is the slope of $\overline{a v_i}$, where $v_i$ is the vertex of the equilateral triangle inside $\bigblacktriangle(a b c)$ with vertices $v_i$, $u_{i}$, and $u_{i+1}$, with $0 \leq i \leq 2\Delta$.

\smallskip
\noindent{\em Magenta slopes:} We have two sets of magenta slopes:

\begin{enumerate}[$\blacktriangleright$]
\item {\em Left-magenta slopes:} The \emph{$i$-th l-magenta slope} $M^l_i$ is $\frac{i \pi}{3 \delta^*}$, with $1 \leq i \leq \delta^*-1$.
For convenience, we let $M^l_{\delta^*} = B_0$ and consider $B_0$ to be also left-magenta.

\item {\em Right-magenta slopes:} The \emph{$i$-th r-magenta} slope $M^r_i$ is $\pi-M^l_i$, with $1 \leq i \leq \delta^*-1$.
Again, we let $M^r_{\delta^*}=B_{2\Delta}$ and consider $B_{2\Delta}$ to be also right-magenta.
\end{enumerate}

\noindent{\em Red slopes:}
Let $M^l_i$ be a left-magenta slope, with $2 \leq i \leq \delta^*$, and let $M^r_j$ be a right-magenta slope, with $2 \leq j \leq \delta^*$.
Also, let $1 \leq h \leq 2\Delta$.
We have:
\smallskip
\begin{enumerate}[$\blacktriangleright$]
\item {\em Central-red slopes:}
Let $\bigblacktriangle(a' b' c')$ be a triangle such that the slope of $\overline{b' c'}$ is the black slope, the slope of $\overline{c' a'}$ is $M^r_j$, and the slope of~$\overline{a' b'}$ is~$M^l_i$.
Let $p^*$ be the intersection point between the line with slope $M^l_{i-1}$ passing through $b'$ and the line with slope $M^l_{j-1}$ passing through $c'$. The \emph{$c$-red slope $R^c_{i,j}$} is the slope of the segment~$\overline{a' p^*}$; see~\cref{fig:main-central}.

\item {\em Left-red slopes:}
Let $q$ be a point above the $x$-axis. Let
$p'$ be the intersection point between the line with slope $M^l_i$ passing through $q$ and the $x$-axis. Also, let $p''$ be the intersection point between the line with slope $O_{h}$ passing through~$q$ and the $x$-axis.
Further, let $p^*$ be the intersection point between the line with slope~$M^l_{i-1}$ passing through $p'$ and the line with slope $B_{2\Delta}$ passing through~$p''$.
The \emph{$l$-red slope $R^{l}_{i,h}$} is the slope of the segment $\overline{q p^*}$; see~\cref{fig:main-left-left}.
\item {\em Right-red slopes:}
Let $q$ be a point above the $x$-axis. Let
$p'$ be the intersection point between the line with slope $O_{h}$ passing through $q$ and the $x$-axis. Also, let~$p''$ be the intersection point between the line with slope $M^r_{j}$ passing through~$q$ and the $x$-axis.
Further, let $p^*$ be the intersection point between the line with slope $B_0$ passing through $p'$ and the line with slope $M^r_{j-1}$ passing through $p''$.
The \emph{$r$-red slope $R^{r}_{h,j}$} is the slope of the segment~$\overline{q p^*}$;~see~\cref{fig:main-left-right}.
\end{enumerate}

\noindent Let $\mathcal{S}$ be the union of these slope sets together with the black slope. Note that, 
\begin{eqnarray}\label{eq:numberOfslopes}
\nonumber|\mathcal{S}|&{=}& 1{+}%
\orange{2\Delta}{+}%
\blue{2\Delta{+}1}{+}%
\purple{2(\delta^*{-}1)}{+}%
\red{(\delta^*{-}1)^2{+}4\Delta(\delta^*{-}1)}\\%
&{=}&\delta^{*2}{+}4\Delta\delta^*{+}1%
{\leq} 5\Delta^2{-}1
\end{eqnarray}

\smallskip\noindent{\bf Construction.}
In what follows we assume that $G$ is rooted at $\rho$, with leftmost path-vertex~$\ell$ and rightmost path-vertex $r$. Further, recall that $\cal T$ is canonical. 
We say that a triangle $\bigblacktriangle(a_\mu b_\mu c_\mu)$ is \emph{good for} a node $\mu$ of $\cal T$, if it satisfies the following properties.
First, the side $\overline{b_\mu c_\mu}$ has the black slope.
Second, the slopes $s_l$ and $s_r$ of the sides $\overline{a_\mu b_\mu}$ and $\overline{a_\mu c_\mu}$, respectively, are such that:
\begin{enumerate}[\bf G.1]
\item \label{prop:both-orange-order} If $s_l = O_i$ and $s_r=O_j$ are orange, then $j=i+1$.
\item \label{prop:s-node-orange-left-magenta} If $\mu$ is an S- or a Q-node, then $s_l$ is either (i) orange or (ii) a left-magenta slope such that $s_l \geq M^l_{\delta_{\mu}(\ell_{\mu})}$;
\item \label{prop:s-node-orange-right-magenta}If $\mu$ is an S- or a Q-node, then $s_r$ is either (i) orange or (ii) a right-magenta slope such that $s_r \leq M^r_{\delta_{\mu}(r_{\mu})}$;
\item \label{prop:good-both-missing} If $\mu$ is an S-node whose pertinent graph contains neither the edge $(\rho_\mu,\ell_\mu)$ nor the edge $(\rho_\mu,r_\mu)$, then at least one among $s_l$ and $s_r$ is an orange slope;
\item \label{prop:p-node-magenta-magenta} If $\mu$ is a P-node, $s_l$
is a left-magenta slope such that $s_l \geq M^l_{\delta_{\mu}(\ell_{\mu})}$ and
$s_r$ is a right-magenta slope such that $s_r \leq M^r_{\delta_{\mu}(r_{\mu})}$.
\end{enumerate}

Let $\pi$ be a planar straight-line drawing of a path $(u_1,\dots,u_k)$ directed from~$u_1$ to~$u_k$, and let $x(u)$ and $y(u)$ denote the $x$- and $y$-coordinate of a vertex $u$, respectively. We say that $\pi$ is \emph{$\rightx$-monotone}, if $y(u_{i+1}) \geq y(u_i)$ and $x(u_{i+1}) > x(u_i)$, for~$i=1,\dots,k-1$. Similarly, we say that it is \emph{$\leftx$-monotone}, if $y(u_{i+1}) \geq y(u_i)$ and~$x(u_{i+1}) < x(u_i)$, for $i=1,\dots,k-1$. Let $\mu$ be a node of $\cal T$ and let $\bigblacktriangle(a_\mu b_\mu c_\mu)$ be a good triangle for $\mu$. Let $s_l$ and~$s_r$ be the slopes of $\overline{a_\mu b_\mu}$ and $\overline{a_\mu c_\mu}$, respectively.
We will recursively construct a planar straight-line drawing $\Gamma_\mu$ of $G_\mu$ with the following {\em geometric properties}. 

\begin{enumerate}[\bf {P.}1]
\item \label{prop:gamma-slopes} $\Gamma_\mu$ uses the slopes in $\mathcal{S}$.
\item \label{prop:gamma-triangle} The convex hull of $\Gamma_\mu$ is the given triangle $\bigblacktriangle (a_\mu b_\mu c_\mu)$, and the vertices $\rho_\mu$, $\ell_\mu$, and~$r_\mu$ are mapped to the points $a_\mu$, $b_\mu$, and~$c_\mu$, respectively. 
\item \label{prop:x-monotonicity} 
If $s_l$ (resp. $s_r$) is left-magenta (resp. right-magenta), then the left path (resp.\ right path) is $\rightx$-monotone (resp. $\leftx$-monotone); if~$s_l$ (resp. $s_r$) is orange, then the left path (resp.\ right path) is $\rightx$-monotone (resp. $\leftx$-monotone) except, possibly, for the edge incident to $\rho_\mu$.
\end{enumerate}
\noindent

 \begin{figure}[t!]
	\centering
	\subfigure[$s_r$ is right-magenta]{\label{fig:orange-magenta}
		\includegraphics[scale=.45,page=14]{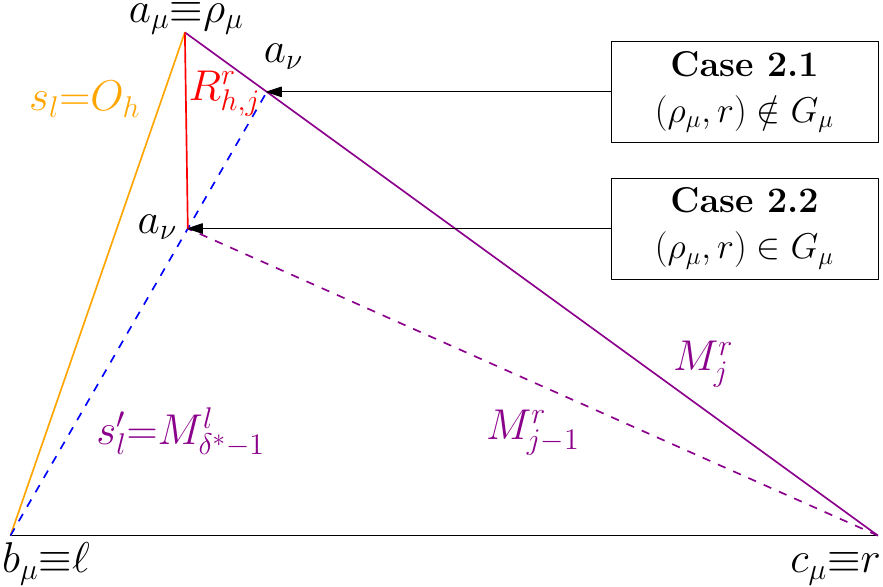}
	}
	\hfil
	\subfigure[$s_r$ is orange]{\label{fig:orange-orange}
		\includegraphics[scale=.45,page=15]{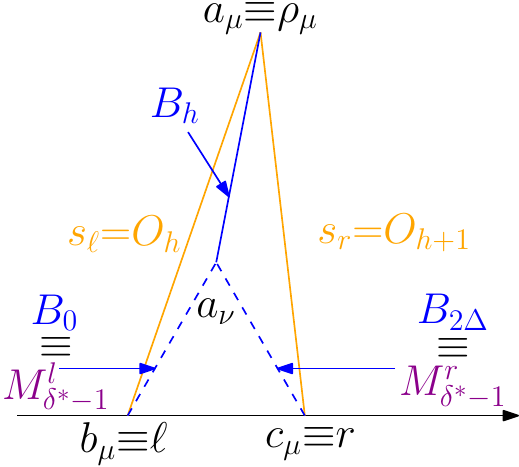}
	}
	\subfigure[\label{fig:magenta-magenta}]{\includegraphics[width=.48\textwidth,page=13]{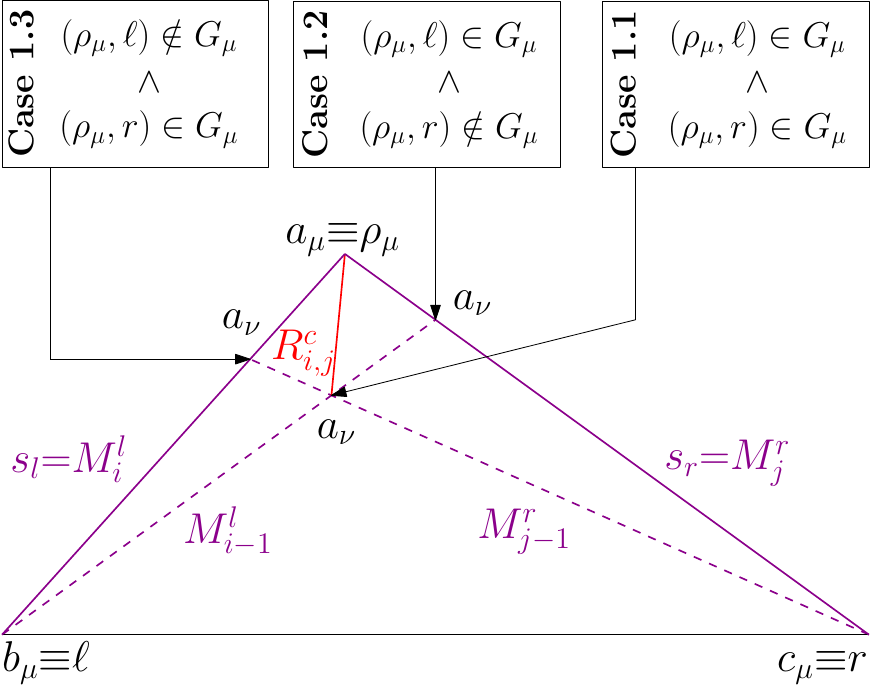}}
	\hfil
	\subfigure[\label{fig:p-node-induction}]{\includegraphics[width=.48\textwidth,page=23]{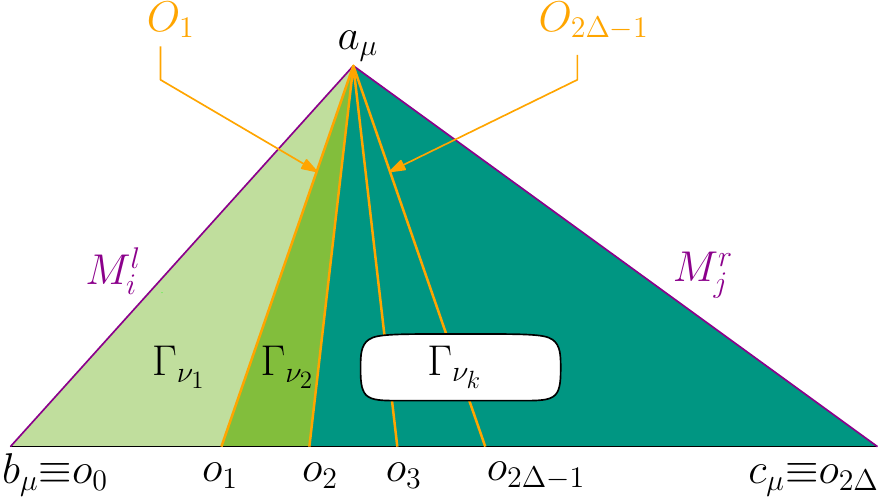}}
		\caption{(a)-(c) Construction of a good triangle for the child of an S-node: (a)-(b) $s_l$ is orange; (c) $s_l$ and $s_r$ are magenta. (d) Construction of good triangles for the children of a P-node with $k=3$ children. The triangle of each child has a distinct opacity.}
\end{figure}

We remark \blue{Property P.}\ref{prop:x-monotonicity} is not needed to compute a drawing of a cycle-tree, but it will turn out to be fundamental to handle~nested~pseudotrees.

\smallskip
\noindent
We describe how to construct $\Gamma_\mu$ in a given good triangle $\bigblacktriangle (a_\mu b_\mu c_\mu)$ for $\mu$, based on the type of $\mu$. When $\mu$ is the root of $\cal T$, the algorithm yields a planar straight-line drawing $\Gamma$ of $G$ using the slopes in $\mathcal S$. 

\smallskip
\noindent\fbox{\sc Q-nodes.} If $\mu$ is a Q-node, we obtain $\Gamma_\mu$ by placing $\rho_\mu$, $\ell_\mu$, and $r_\mu$ at the points~$a_\mu$,~$b_\mu$, and $c_\mu$, respectively. %

\smallskip

\noindent\fbox{\sc S-nodes.} If $\mu$ is an S-node, then the construction of $\Gamma_\mu$ depends on the degree of~$\rho_\mu$ in $G_\mu$. Let $\nu$ be the unique child of $\mu$.
For convenience, we let $\ell = \ell_\mu = \ell_\nu$ and~$r = r_\mu = r_\nu$.
We first recursively build a drawing $\Gamma_\nu$ of $G_\nu$ in a triangle~$\bigblacktriangle (a_\nu b_\nu c_\nu)$ that is good for $\nu$, where $a_\nu$ is appropriately placed in the interior of $\bigblacktriangle (a_\mu b_\mu c_\mu)$ while $b_\nu=b_\mu$ and $c_\nu=c_\mu$.
Then, $\Gamma_\mu$ is obtained from $\Gamma_\nu$ by simply placing $\rho_\mu$ at $a_\mu$, and by drawing the edges incident to $\rho_\mu$ as straight-line segments.

 Note that, $\rho_\mu$ is adjacent to the root $\rho_\nu$
 of $G_\nu$, and to either $\ell$, or $r$, or both.
 In order to define the point $a_\nu$, we now choose the slopes $s'_l$ and $s'_r$ of the segments~$\overline{a_\nu b_\nu}$ and $\overline{a_\nu c_\nu}$, respectively, as follows.
 We start with~$s'_l$.
 Since $\mu$ is an S-node, $s_l$ is either orange or a left-magenta slope $M^l_i$.
 If $s_l$ is orange, then $s'_l = M^l_{\delta^*}$. See \Cref{fig:orange-magenta} and~\Cref{fig:orange-orange}.
 If $s_l=M^l_i$ and $(\rho_\mu,\ell)$ belongs to $G_\mu$, we have that $s'_l=M^l_{i-1}$. Notice that, by \blue{Property G.}\ref{prop:s-node-orange-left-magenta} $i \geq \delta_{\mu}(\ell)$, and since $\ell$ is incident at least to $(\rho_\mu,\ell)$ and to an edge of the path of $G$, we have $i \geq 2$. If $s_l=M^l_i$ and $(\rho_\mu,\ell)$ does not belong to $G_\mu$, we have that $s'_l = s_l = M^l_i$. See \Cref{fig:magenta-magenta}.
 The choice of $s'_r$ is symmetric, based on the existence of $(\rho_\mu,r)$. Notice that, by \blue{Property G.}\ref{prop:good-both-missing}, if both~$s_l$ and $s_r$ are magenta, then one between $(\rho_\mu,\ell)$ and  $(\rho_\mu,r)$ exists.

\smallskip

\noindent\fbox{\sc P-nodes.}  If $\mu$ is a P-node, then let $\nu_1,\nu_2,\dots,\nu_k$, with $2 \leq k \leq 2\Delta+1$ be the children of $\mu$. Since $\mathcal T$ is canonical, no $\nu_i$ is a P-node.
 Refer to \cref{fig:p-node-induction}.
Let $o_i$ be the intersection point between $\overline{b_\mu c_\mu}$ and the line passing through $a_\mu$ with slope $O_i$, for $i=1,\dots,2\Delta-1$. For convenience, we let $o_0 = b_\mu$ and $o_{2\Delta}=c_\mu$.
We recursively build a drawing $\Gamma_{\nu_i}$ of $G_{\nu_i}$, with $i=1,\dots,k-1$, in the triangle~$\bigblacktriangle (a_\mu o_{i-1} o_i)$, which is good for $\nu_i$, and a drawing $\Gamma_{\nu_k}$ of $G_{\nu_k}$ in the triangle~$\bigblacktriangle (a_\mu o_{k-1} o_{2\Delta})$, which is good for $\nu_k$. $\Gamma_\mu$ is the union of~the~$\Gamma_{\nu_i}$'s.

\medskip 
\noindent \textbf{Proof of correctness.} We now show that the construction satisfies \blue{Properties P.}\ref{prop:gamma-slopes}, \blue{P.}\ref{prop:gamma-triangle}, and \blue{P.}\ref{prop:x-monotonicity}.

The fact that \blue{Properties P.}\ref{prop:gamma-slopes}, \blue{P.}\ref{prop:gamma-triangle}, and \blue{P.}\ref{prop:x-monotonicity} are satisfied by the drawing $\Gamma_\mu$ when~$\mu$ is a Q-node trivially follows by construction. Hence, it remains to consider~S- and P-nodes.

Recall that the degree of a vertex $v$ in $G_\mu$ is denoted by $\deg_{G_\mu}(v)$; since this leads no confusion here,  we let $\deg_\mu(v)=\deg_{G_\mu}(v)$.

\begin{lemma}\label{lem:s-nodes}
	Let $\mu$ be an S-node. $\Gamma_\mu$ satisfies \blue{Properties P.}\ref{prop:gamma-slopes}, \blue{P.}\ref{prop:gamma-triangle}, and \blue{P.}\ref{prop:x-monotonicity}.
\end{lemma}

\begin{proof}
	For each of the cases in the construction of \Cref{ss:3connected}, we start by proving the following:
	\begin{inparaenum}[(i)]
		\item the triangle $\bigblacktriangle (a_\nu b_\mu c_\mu)$ is good for $\nu$,
		\item the slope of $\overline{a_\nu a_\mu}$ belongs~to~$\cal S$,
		and
		\item $\Gamma_\mu$ satisfies \blue{Property P.}\ref{prop:x-monotonicity}.
	\end{inparaenum}
	Then, we prove that $\Gamma_\mu$ is a planar straight-line drawing, and that it satisfies \blue{Properties P.}\ref{prop:gamma-slopes} and \blue{P.}\ref{prop:gamma-triangle}.
	
	Observe that, by construction, neither $s'_l$ nor $s'_r$ are orange. Thus, $\bigblacktriangle (a_\nu b_\mu c_\mu)$ trivially satisfies \blue{Property G.}\ref{prop:both-orange-order}. Furthermore, since $G_\mu$ is an almost-$3$-connected path-tree and since $\mu$ is an S-node, we have that if $\nu$ is also an $S$-node, then at least one of the edges $(\rho_\nu,\ell)$ and $(\rho_\nu,r)$ must exist. Thus, $\bigblacktriangle (a_\nu b_\mu c_\mu)$ also trivially satisfies \blue{Property G.}\ref{prop:good-both-missing}. Due to these observations, in order to prove (i), it remains to argue about \blue{Properties G.}\ref{prop:s-node-orange-left-magenta},  \blue{G.}\ref{prop:s-node-orange-right-magenta}, and \blue{G.}\ref{prop:p-node-magenta-magenta}.
	The proof splits into four cases, corresponding to the four possible combinations of colors for the slopes of~$s_l$ and~$s_r$. In all cases, we show that~$s'_l$ (resp. $s'_r$) is a left-magenta slope (resp.\ right-magenta slope) which satisfies \blue{Properties G.}\ref{prop:s-node-orange-left-magenta} and \blue{G.}\ref{prop:p-node-magenta-magenta} (resp.~\blue{Properties~G.}\ref{prop:s-node-orange-right-magenta}~and~\blue{G.}\ref{prop:p-node-magenta-magenta}). Along the way, we also prove (ii) and (iii). We start by observing that if $s_l$ is a left-magenta slope $M^l_i$ and the edge $(\rho_\mu,\ell)$  belongs to $G_{\mu}$, then $i \geq 2$. Namely, by \blue{Property G.}\ref{prop:s-node-orange-left-magenta} $i \geq \delta_{\mu}(\ell)$, and since $\ell$ is incident to an edge of the path of $G$, if $(\rho_\mu,\ell)$ belongs to $G_{\mu}$, then $\delta_{\mu}(\ell) \geq 2$. Analogously, if $s_r$ is a right-magenta slope $M^r_j$ and the edge $(\rho_\mu,r)$  belongs to $G_{\mu}$, then $j \geq 2$.

	\smallskip
	{\bf Case~1:} Both $s_l$ and $s_r$ are magenta slopes, i.e., $s_l = M^l_i$ and $s_r = M^r_j$, with~$1 \leq i \leq \delta^*$ and $1 \leq j \leq \delta^*$. Since $\bigblacktriangle (a_\mu b_\mu c_\mu)$ is good for $\mu$, \blue{Property G.}\ref{prop:good-both-missing} implies that at least one of $(\rho_\mu,\ell)$ and $(\rho_\mu,r)$ belongs to $G_\mu$.  There are three subcases Case~1.1, Case~1.2, and Case~1.3. Refer to \cref{fig:magenta-magenta-app}.

	\begin{figure}[h!]
		\centering
		\includegraphics[width=.5\textwidth,page=13]{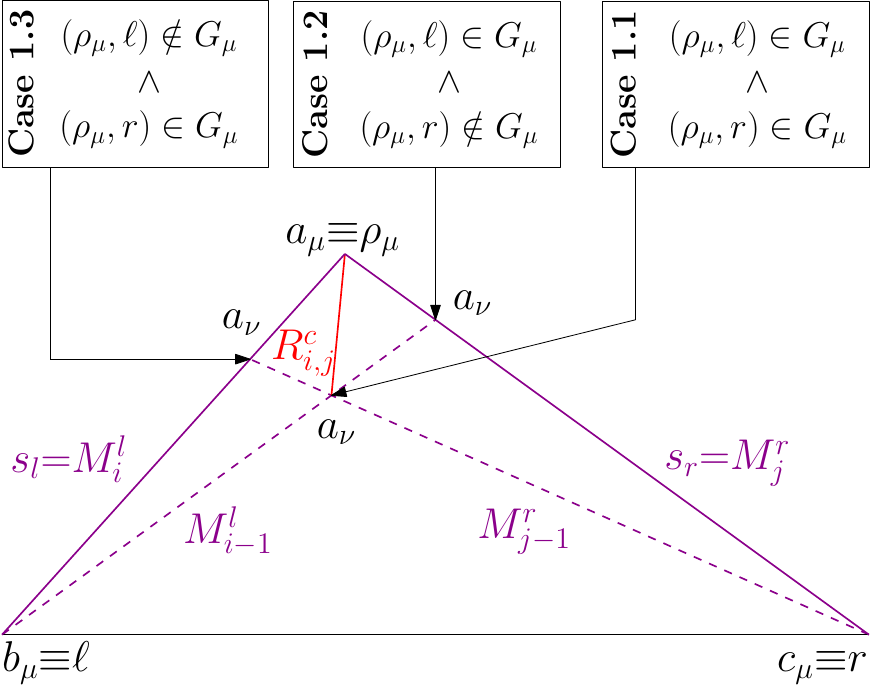}
		\caption{Construction of triangle $\bigblacktriangle (a' b c)$ when both $s_l$ and $s_r$ are magenta.}
		\label{fig:magenta-magenta-app}
	\end{figure}
	
	Case 1.1: both $(\rho_\mu,\ell)$ and $(\rho_\mu,r)$ belong to $G_\mu$. In this case, $s'_l = M^l_{i-1}$ and~$s'_r = M^r_{j-1}$. Since $i,j \geq 2$ both $M^l_{i-1}$ and $M^r_{j-1}$ exist. We have $\deg_{\nu}(\ell)=\deg_{\mu}(\ell)-1$ and $\deg_{\nu}(r)=\deg_{\mu}(r)-1$. Therefore, $s'_l = M^l_{i-1} \geq M^l_{\delta_{\nu}(\ell)}$, since $s_l = M^l_{i} \geq M^l_{\delta_{\mu}(\ell)}$, and $s'_r = M^r_{j-1} \leq M^r_{\delta_{\nu}(r)}$, since $s_r = M^r_{j} \leq M^r_{\delta_{\mu}(r)}$. It follows that
	$\bigblacktriangle (a_\nu b_\mu c_\mu)$ is good~for~$\nu$.
	The slope of the segment $\overline{a_\nu a_\mu}$ is the $c$-red slope $R^c_{i,j}$.
	Furthermore, both the left and the right path of $\mu$ consist of the edge $(\rho_\mu,\ell_\mu)$ and the edge~$(\rho_\mu,r_\mu)$, respectively, and thus  $\Gamma_\mu$ trivially satisfies \blue{Property P.}\ref{prop:x-monotonicity}.
	
	Case 1.2: $(\rho_\mu,\ell)$ belongs to $G_\mu$ and $(\rho_\mu,r)$ does not belong to $G_\mu$. In this case,~$s'_l = M^l_{i-1}$ and $s'_r = M^r_{j}$. Since $i \geq 2$, $M^l_{i-1}$ exists.
	We have~$\deg_{\nu}(\ell)=\deg_{\mu}(\ell)-1$ and $\deg_{\nu}(r)=\deg_{\mu}(r)$. Similarly to Case~1.1, $s'_l = M^l_{i-1} \geq M^l_{\delta_{\nu}(\ell)}$, and $s'_r = M^r_{j} \leq M^r_{\delta_{\nu}(r)}$. It follows that $\bigblacktriangle (a_\nu b_\mu c_\mu)$ is good~for~$\nu$.
	The slope of the segment $\overline{a_\nu a_\mu}$ is the $r$-magenta slope $M^r_j$.
	Furthermore, the left path of $\mu$ consists of the edge $(\rho_\mu,\ell)$. Also, the right path of $\mu$ consists of the right path $\pi_r$ of $\nu$ and the edge $(\rho_\nu,\rho_\mu)$. Since $s'_r$ is a right-magenta slope, by \blue{Property P.}\ref{prop:x-monotonicity} of $\Gamma_\nu$, we have that $\pi_r$ is $\leftx$-monotone. Finally, since the slope of the edge
	$(\rho_\nu,\rho_\mu)$ is $M^r_j$, we have that $\rho_\mu$ lies above and to the left of $\rho_\nu$. Therefore,  $\Gamma_\mu$ satisfies \blue{Property P.}\ref{prop:x-monotonicity}.

	Case~1.3: $(\rho_\mu,\ell)$ does not belong to $G_\mu$ and $(\rho_\mu,r)$ belongs to $G_\mu$.
	We have $\deg_{\nu}(\ell)=\deg_{\mu}(\ell)$ and $\deg_{\nu}(r)=\deg_{\mu}(r)-1$. We have $s'_l = M^l_{i} \geq M^l_{\delta_{\nu}(\ell)}$, and $s'_r = M^r_{j-1} \leq M^r_{\delta_{\nu}(r)}$. Since $j \geq 2$, $M^r_{j-1}$ exists. It follows that $\bigblacktriangle (a_\nu b_\mu c_\mu)$ is good~for~$\nu$.
	The slope of the segment $\overline{a_\nu a_\mu}$ is the $l$-magenta slope $M^l_i$.
	Furthermore, the right path of $\mu$ consists of the edge $(\rho_\mu,r)$. Also, the left path of $\mu$ consists of the left path $\pi_\ell$ of $\nu$ and the edge $(\rho_\nu,\rho_\mu)$. Since $s'_\ell$ is a left-magenta slope, by \blue{Property P.}\ref{prop:x-monotonicity} of $\Gamma_\nu$, we have that $\pi_\ell$ is $\rightx$-monotone. Finally, since the slope of the edge
	$(\rho_\nu,\rho_\mu)$ is $M^l_i$, we have that $\rho_\mu$ lies above and to the right of $\rho_\nu$. Therefore,~$\Gamma_\mu$ satisfies~\blue{Property~P.}\ref{prop:x-monotonicity}.
	
	\smallskip
	{\bf Case~2:} The slopes $s_l$ and $s_r$ are orange and right-magenta, respectively. That is, $s_l = O_h$, with $1 \leq h \leq 2\Delta$, and $s_r = M^r_j$, with $1 \leq j \leq \delta^*$.
	Note that, by construction, $s'_l = M^l_{\delta^*}$.
	We distinguish two subcases, based on whether $(\rho_\mu,r)$ belongs to $G_\mu$. Refer to \cref{fig:orange-magenta-app}.
	
	\begin{figure}[h!]
		\centering
		\includegraphics[width=.5\textwidth,page=14]{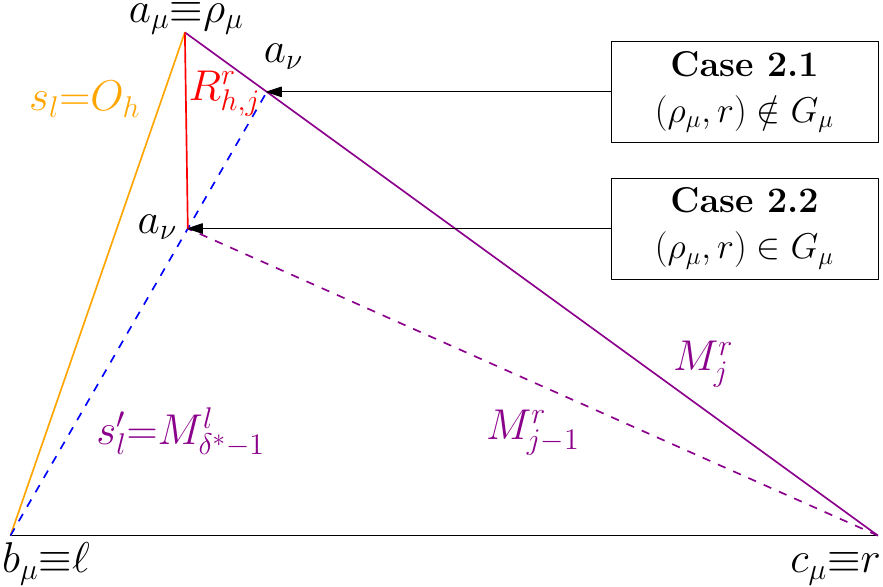}
		\caption{Construction of triangle $\bigblacktriangle (a_\nu b_\mu c_\mu)$ when $s_l$ is orange and $s_r$ is right-magenta.}
		\label{fig:orange-magenta-app}
	\end{figure}

	Case~2.1: $(\rho_\mu,r)$ does not belong to $G_\mu$. In this case, $s'_r = M^r_{j} = s_r$. We have that $\deg_\nu(\ell) \leq \deg_\mu(\ell)$ and $\deg_\nu(r) = \deg_\mu(r)$. Therefore, $s'_l=M^l_{\delta^*} \geq M^l_{\delta_\nu(\ell)}$, since~$M^l_{\delta^*}$ is the largest left-magenta slope, and $s'_r =  M^r_{j} \leq M^r_{\delta_\nu(r)}$, since $s_r = M^r_{j} \leq M^r_{\delta_\mu(r)}$. It follows that $\bigblacktriangle (a_\nu b_\mu c_\mu)$ is good~for~$\nu$.
	The slope of the segment~$\overline{a_\nu a_\mu}$ is the $r$-magenta slope $M^r_j$.
	The proof that $\Gamma_\mu$ satisfies \blue{Property P.}\ref{prop:x-monotonicity} is the same as in Case 1.2.

	Case~2.2: $(\rho_\mu,r)$ belongs to $G_\mu$. In this case, $s'_r = M^r_{j-1}$. Since $j \geq 2$, $M^r_{j-1}$ exists. We have that $\deg_\nu(\ell) \leq \deg_\mu(\ell)$ and $\deg_\nu(r) = \deg_\mu(r)-1$. Therefore, $s'_l=M^l_{\delta^*} \geq M^l_{\delta_\nu(\ell)}$ as for Case~2.1, and $s'_r =  M^r_{j-1} \leq M^r_{\delta_\nu(r)}$, since $s_r = M^r_{j} \leq M^r_{\delta_\mu(r)}$. It follows that $\bigblacktriangle (a_\nu b_\mu c_\mu)$ is good~for~$\nu$.
	The slope of the segment $\overline{a_\nu a_\mu}$ is the $r$-red slope $R^r_{h,j}$.
	The proof that $\Gamma_\mu$ satisfies \blue{Property P.}\ref{prop:x-monotonicity} is the same as in Case 1.1, if~$(\rho_\mu,\ell)$ belongs to $G_\mu$. Otherwise, the right path of $\mu$ consists of the edge $(\rho_\mu,r)$. Also, the left path of $\mu$ consists of the left path $\pi_\ell$ of $\nu$ and the edge~$(\rho_\nu,\rho_\mu)$. Since~$s'_\ell$ is a left-magenta slope, by \blue{Property P.}\ref{prop:x-monotonicity} of $\Gamma_\nu$, we have that~$\pi_\ell$ is $\rightx$-monotone. Therefore, $\Gamma_\mu$ satisfies \blue{Property P.}\ref{prop:x-monotonicity}.

	\smallskip
	{\bf Case~3:} The slopes $s_l$ and $s_r$ are left-magenta and orange, respectively. I.e.,~$s_l = M^l_i$, with $1 \leq i \leq \delta^*$, and $s_r = O_h$, with $1 \leq h \leq 2\Delta$.
	The proof of this case  is based on two subcases symmetric to those of Case~2. Namely, Case~3.1 (i.e., $(\rho_\mu,r)$ does not belong to $G_\mu$) and Case~3.2 (i.e., $(\rho_\mu,r)$ belongs to $G_\mu$).
	In particular, the slope of the segment $\overline{a_\nu a_\mu}$ is the $l$-magenta slope $M^l_i$ in Case~3.1 and the $l$-red slope $R^l_{i,h}$ in Case~3.2. The proof that $\Gamma_\mu$ satisfies \blue{Property P.}\ref{prop:x-monotonicity} is also symmetric to the one in Case~2.

	\begin{figure}[h!]
		\centering
		\includegraphics[width=.4\textwidth,page=15]{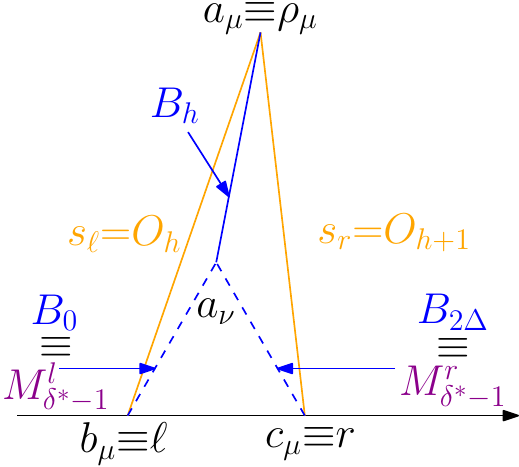}
		\caption{Construction of triangle $\bigblacktriangle (a' b c)$ when $s_l$ and $s_r$ are orange.}
		\label{fig:orange-orange-app}
	\end{figure}
	
	\smallskip
	{\bf Case~4:} The slopes $s_l$ and $s_r$ are orange. By \blue{Property~G.}\ref{prop:both-orange-order}, we have that $s_l = O_h$ and $s_r=O_{h+1}$, with $1 \leq h \leq 2\Delta$.
	Refer to \cref{fig:orange-orange-app}.
	In this case, $s'_l = M^l_{\delta^*}$ and $s'_r=M^r_{\delta^*}$.
	We have that $\deg_\nu(\ell) \leq \deg_\mu(\ell)$ and $\deg_\nu(r) \leq \deg_\mu(r)$.
	Therefore, $s'_l=M^l_{\delta^*} \geq M^l_{\delta_\nu(\ell)}$, since $M^l_{\delta^*}$ is the largest left-magenta slope, and $s'_r=M^r_{\delta^*} \leq M^l_{\delta_\nu(\ell)}$, since $M^r_{\delta^*}$ is the smallest right-magenta slope. It follows that $\bigblacktriangle (a_\nu b_\mu c_\mu)$ is good~for~$\nu$.
	The slope of the segment $\overline{a_\nu a_\mu}$ is the blue slope slope $B_{h}$.
	Finally, we argue about \blue{Property P.}\ref{prop:x-monotonicity} of $\Gamma_\mu$.
	We only consider the left path of $\mu$, as the right path can be treated symmetrically.
	Recall that, since $s_\ell$ is orange, in order to satisfy this property, the left path of $\mu$ needs to be $\rightx$-monotone, except for its edge incident to $\rho_\mu$.
	If the edge $(\rho_\mu,\ell)$ belongs to $G_\mu$, then the left path of $\mu$ consists of just the edge $(\rho_\mu,\ell)$.
	Otherwise, the left path of $\mu$ consists of the left path $\pi_\ell$ of $\nu$ and the edge $(\rho_\nu,\rho_\mu)$. Since $s'_\ell$ is a left-magenta slope, by \blue{Property P.}\ref{prop:x-monotonicity} of $\Gamma_\nu$, we have that $\pi_\ell$ is $\rightx$-monotone.
	This proves that~$\Gamma_\mu$ satisfies \blue{Property P.}\ref{prop:x-monotonicity}, which concludes the proof of (i), (ii),~and~(iii).
	
	It remains to prove that $\Gamma_\mu$ is a planar straight-line drawing, and that it satisfies \blue{Properties P.}\ref{prop:gamma-slopes} and \blue{P.}\ref{prop:gamma-triangle}.

	First, since $G_\nu$ contains one vertex less than $G_\mu$ (namely, the root~$\rho_\mu$ of~$\mu$), the planar straight-line drawing $\Gamma_\nu$ of $G_\nu$ can be recursively constructed in $\bigblacktriangle(a_\nu b_\mu c_\mu)$ so to satisfy
	\blue{Properties P.}\ref{prop:gamma-slopes}, \blue{P.}\ref{prop:gamma-triangle}, and \blue{P.}\ref{prop:x-monotonicity}.
	
	Second, in all the cases described above, as depicted in \cref{fig:magenta-magenta,fig:orange-magenta,fig:orange-orange}, the point $a_\nu$ lies either on the line-segment $\overline{a_\mu b_\mu}$, or on the line-segment of $\overline{a_\mu c_\mu}$, or in the interior of $\bigblacktriangle(a_\mu b_\mu c_\mu)$ by construction. Thus, $\bigblacktriangle(a_\nu b_\mu c_\mu)$ is contained in~$\bigblacktriangle(a_\mu b_\mu c_\mu)$.
	Furthermore, the placement of $a_\nu$ is such that it is possible to draw the edges incident to $\rho_\mu$ in $G_\mu$ as straight-line segments $\overline{a_\nu a_\mu}$, $\overline{a_\nu b_\mu}$, and $\overline{a_\nu c_\mu}$ that do not cross $\bigblacktriangle(a_\nu b_\mu c_\mu)$, except at its corners. Finally, as already shown, the slopes of these segments belong~to~$\cal S$.\qed\end{proof}

\begin{lemma}\label{lem:p-nodes}
	Let $\mu$ be a P-node. $\Gamma_\mu$ satisfies \blue{Properties P.}\ref{prop:gamma-slopes}, \blue{P.}\ref{prop:gamma-triangle}, and \blue{P.}\ref{prop:x-monotonicity}.
\end{lemma}

\begin{proof}
	First, we prove that the triangles defined above are good for the respective child of $\mu$. Then, we prove that $\Gamma_\mu$ is a planar straight-line drawing that satisfies \blue{Properties P.}\ref{prop:gamma-slopes}, \blue{P.}\ref{prop:gamma-triangle}, and \blue{P.}\ref{prop:x-monotonicity}.
	
	First, observe that since $\T$ is canonical no $\nu_i$ is a P-node and therefore  \blue{Property G.}\ref{prop:p-node-magenta-magenta} is trivially satisfied by all the defined triangles.
	
	For $i=2,\dots,k-1$, consider the triangle $\bigblacktriangle (a_\mu o_{i-1}o_i)$. We have that the slopes of $\overline{a_\mu  o_{i-1}}$ and $\overline{a_\mu o_{i}}$ are the orange slopes $O_{i-1}$ and $O_{i}$, respectively. Therefore, \blue{Properties G.}\ref{prop:both-orange-order}, \blue{G.}\ref{prop:s-node-orange-left-magenta}, \blue{G.}\ref{prop:s-node-orange-right-magenta}, and \blue{G.}\ref{prop:good-both-missing} are satisfied.
	It follows that $\bigblacktriangle (a_\mu o_{i-1}o_i)$ is good for $\nu_i$, for $i=2,\dots,k-1$.

	If $i=1$, consider the triangle $\bigblacktriangle (a_\mu o_{0} o_1)$.
	We have that the slope of $\overline{a o_{0}}$ is~$s_l \geq M^l_{\delta(\ell)}$, since $\bigblacktriangle (a_\mu b_\mu c_\mu)$ satisfies \blue{Property G.}\ref{prop:p-node-magenta-magenta} as it is good for $\mu$.
	Therefore, \blue{Property G.}\ref{prop:both-orange-order} is trivially satisfied, since $s_l$ is not orange,
	and \blue{Property G.}\ref{prop:s-node-orange-left-magenta} is satisfied, since $\deg_{\nu_1}(\ell)=\deg_{\mu}(\ell)$.
	Furthermore, the slope of $\overline{a_\mu o_{1}}$ is the orange slope~$O_{1}$.
	Therefore, \blue{Properties G.}\ref{prop:s-node-orange-right-magenta} and \blue{G.}\ref{prop:good-both-missing} are satisfied. It follows that $\bigblacktriangle (a_\mu o_{0} o_1)$ is good for $\nu_1$.

	If $i=k$, consider the triangle $\bigblacktriangle (a_\mu o_{k-1} o_{2\Delta})$.
	We have that the slope of $\overline{a_\mu o_{k-1}}$ is the orange slope $O_{k-1}$.
	Therefore, \blue{Properties G.}\ref{prop:s-node-orange-left-magenta} and \blue{G.}\ref{prop:good-both-missing} are satisfied.
	Furthermore, we have that the slope of $\overline{a_\mu o_{2\Delta}}$ is $s_r \leq M^r_{\delta(r)}$, since $\bigblacktriangle (a_\mu b_\mu c_\mu)$ satisfies \blue{Property G.}\ref{prop:p-node-magenta-magenta} as it is good for $\mu$.
	Therefore, \blue{Property G.}\ref{prop:both-orange-order} is trivially satisfied, since~$s_r$ is not orange,
	and \blue{Property G.}\ref{prop:s-node-orange-right-magenta} is satisfied, since $\deg_{\nu_k}(r)=\deg_{\mu}(r)$.
	It follows that $\bigblacktriangle (a_\mu u_{k-1} u_{2\Delta})$ is good for $\nu_k$.
	This concludes the proof that  each triangle $\bigblacktriangle (a_\mu u_{i-1}u_i)$ is good for $\nu_i$, for $i=1,\dots,k$.
	
	It remains to prove that $\Gamma_\mu$ is a planar straight-line drawing that satisfies \blue{Properties P.}\ref{prop:gamma-slopes}, \blue{P.}\ref{prop:gamma-triangle}, and \blue{P.}\ref{prop:x-monotonicity}.
	
	First, the pertinent graph $G_{\nu_i}$ of each child $\nu_i$, for $i=1,\dots,k$, contains at least one vertex less than $G_\mu$. In fact, the vertex set of each pertinent graph $G_{\nu_i}$ contains at least three vertices and shares with any other pertinent graph $G_{\nu_j}$, with~$j \neq i$, the root $\rho_\mu$ and at most one cycle-vertex. Therefore, the planar straight-line drawing $\Gamma_{\nu_i}$ of each child $\nu_i$, for $i=1,\dots,k$ can be recursively constructed in the respective good triangle so to satisfy
	\blue{Properties P.}\ref{prop:gamma-slopes}, \blue{P.}\ref{prop:gamma-triangle},~and~\blue{P.}\ref{prop:x-monotonicity}.
	
	Second, observe that the triangles defined for the children of $\mu$ are all internally disjoint.
	Thus, $\Gamma_\mu$ is a planar straight-line drawing of $G_\mu$, given that each drawing~$\Gamma_{\nu_i}$ is straight-line and planar. Finally, $\Gamma_\mu$ satisfies~\blue{Properties~P.}\ref{prop:gamma-slopes}~and~\blue{P.}\ref{prop:gamma-triangle} due to the fact that each drawing $\Gamma_{\nu_i}$ satisfies the same properties, and satisfies \blue{Properties~P.}\ref{prop:x-monotonicity} since, in particular, $\Gamma_{\nu_1}$ and $\Gamma_{\nu_k}$ satisfy this property.
\qed\end{proof}

\smallskip
\noindent The following lemma summarizes the results of this section.

\begin{lemma}\label{lem:psn-almost-$3$-connected}
For any almost-$3$-connected path-tree $G$ and
any triangle $\bigblacktriangle (a b c)$ that is good for the root of an SPQ-tree of $G$, 
the graph $G$
admits an embedding-preserving  planar straight-line drawing inside $\bigblacktriangle (a b c)$ that satisfies \blue{Properties P.}\ref{prop:gamma-slopes}, \blue{P.}\ref{prop:gamma-triangle}, and \blue{P.}\ref{prop:x-monotonicity}.
\end{lemma}

\noindent We conclude with two remarks concerning the allocation of slopes.

\begin{remark}\label{remark:red-slope}
	The slope of an edge incident to a path vertex is either orange, magenta, or blue, and in particular it is not red.
\end{remark}

The next remark is a consequence of the fact that red slopes are only used when constructing a drawing of an S-node $\mu$ with child $\nu$. In this case, at most one red slope is used inside the good triangle $\bigblacktriangle (a_\mu b_\mu c_\mu)$ to connect $\rho_{\mu}$ with $\rho_{\nu}$. Since none of the sides of $\bigblacktriangle (a_\mu b_\mu c_\mu)$ uses a red slope, we have the following. 

\begin{remark}\label{remark:non-consecutive-red-slope}
    Let $e_1$ and $e_2$ be two edges having red slopes that are consecutive in the counterclockwise circular order around a common vertex $v$. Let $r_i$ be the ray originating at $v$ and containing the edge $e_i$, for $i=1,2$, and let $W$ be any of the two wedges defined by $r_1$ and $r_2$. There exists a non-red slope $s \in \mathcal S$ such that the ray originating at $v$ with slope $s$ lies inside $W$.
\end{remark}

\begin{figure}[h]
	\centering
	\includegraphics[scale=1.2,page=22]{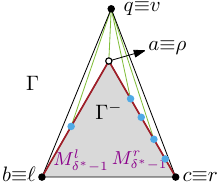}
	\caption{How to draw a \mbox{$3$-connected} cycle-tree.}
	\label{fig:$3$-connected-cycle-tree}
\end{figure}

\smallskip\noindent{\bf $3$-connected cycle-trees.} Let $G$ be a degree-$\Delta$ $3$-connected cycle-tree. We show how to exploit \cref{lem:psn-almost-$3$-connected} to draw $G$ using $O(|{\cal S}|)=O(\Delta^2)$ slopes. Similarly to path-trees, we call \emph{cycle-vertices} the vertices on the outer boundary of $G$ and \emph{tree-vertices} the remaining vertices of $G$.
Let $\ell$, $v$, and $r$ be three cycle-vertices that appear in this clockwise order along the outer face of~$G$; refer to \cref{fig:$3$-connected-cycle-tree}.
Remove~$v$ and its incident edges from~$G$. Denote by~$G^-$ the resulting topological graph. Let~$\pi$ be the graph formed by the edges that belong to the outer face of~$G^-$ and do not belong to the outer face of~$G$.

Since~$G$ is $3$-connected, we have that $G^-$ is at least $2$-connected and that $\pi$ is a path connecting~$\ell$ and $r$ that contains at least one tree-vertex different from $v$. Let~$\rho$ be {\em any} such vertex encountered when traversing $\pi$ from $\ell$ to $r$.
Moreover, the only degree-$2$ vertices of $G^-$, if any, belong to~$\pi$. Let~$G^*$ be the graph obtained from $G^-$ by replacing each degree-$2$ vertex of~$\pi$ different from~$\ell$, $\rho$, and $r$, if any, with an edge connecting its endpoints.
Graph $G^*$ is an almost-$3$-connected path-tree rooted at~$\rho$, with leftmost path-vertex~$\ell$ and rightmost path-vertex $r$.

\begin{lemma}\label{lem:psn-triconnected-cycle-trees-graph}
Every $3$-connected cycle-tree $G$ with maximum degree $\Delta$ has $\psn(G) \in O(|{\cal S}|)$.
\end{lemma}

\begin{proof}
If the outer boundary of $G$ has $3$ vertices, the total number of edges of $G$ is $O(\Delta)$ and hence $\psn(G) \in O(\Delta) \subseteq O(|\cal S|)$. So assume that the outer boundary of~$G$ has more than $3$ vertices.

Let $\cal T$ be the SPQ-tree of $G^*$ and let $\bigblacktriangle(a b c)$ be an equilateral triangle.
Note that an equilateral triangle is good for the root of $\cal T$, regardless of its type. 
Let $\Gamma^*$ be the planar straight-line drawing of $G^*$ inside $\bigblacktriangle(a b c)$, obtained by applying \cref{lem:psn-almost-$3$-connected}. 
We prove that there exists a planar straight-line drawing $\Gamma$ of $G$ such that $\psn(\Gamma) \leq \psn(\Gamma^*) + \Delta$, which implies the statement because $\psn(\Gamma^*) \in O(|{\cal S}|)=O(\Delta^2)$ by \cref{lem:psn-almost-$3$-connected}.
Note that, the slopes $s_\ell$ and $s_r$ of $\overline{ab}$ and $\overline{ac}$ are the largest $l$-magenta slope $M^l_{\delta^*}$ and the smallest $r$-magenta slope $M^r_{\delta^*}$, respectively.
Moreover, since the drawing $\Gamma^*$ inside $\bigblacktriangle(a b c)$ has been obtained by applying \cref{lem:psn-almost-$3$-connected}, we have that $\Gamma^*$ satisfies \blue{Property P.}\ref{prop:x-monotonicity}. %
 We construct a planar straight-line drawing $\Gamma$ of~$G$ as follows; refer~to~\cref{fig:$3$-connected-cycle-tree}. First, we obtain a planar straight-line drawing $\Gamma^-$ of $G^-$ from~$\Gamma^*$, by subdividing the edges that stemmed from the contraction operations (which yielded~$G^*$ from~$G^-$). Clearly, $\psn(\Gamma^-) = \psn(\Gamma^*)$.
$\Gamma^-$ exhibits the following useful property:
By \blue{Property P.}\ref{prop:x-monotonicity} of $\Gamma^*$, we have that the subpath of~$\pi$ from $\ell$ to $\rho$ is $\rightx$-monotone and that the subpath of~$\pi$ from $r$ to~$\rho$ is $\leftx$-monotone.
Second, we select a point $q$ vertically above $\rho$ such that all the straight-line segments connecting $q$ to each of the vertices of $\pi$ do not cross~$\Gamma^-$.
The existence of such a point is guaranteed by the above property.
Finally, we obtain~$\Gamma$ from $\Gamma^-$ by placing $v$ at point $q$, and by drawing its incident edges as straight-line segments. Since $v$ has at most degree $\Delta$, we have that $\psn(\Gamma) \leq \psn(\Gamma^-) + \Delta$.
\qed\end{proof}

By \Cref{remark:red-slope} and since the slopes of the edges incident to $q$ do not belong to the set $\mathcal S$,  we have the following

\begin{remark}\label{rem:red-slope-cycle}
	The slope of an edge incident to a cycle vertex is not red.
\end{remark}

In the next sections, we extend the result of \cref{lem:psn-triconnected-cycle-trees-graph} to $2$-connected and then $1$-connected graphs.

\subsection{$2$-Connected Cycle-Tree Graphs}

Throughout this section $G$ is a $2$-connected cycle-tree. 
We can assume that $G$ is not series-parallel because otherwise it can be drawn with $O(\Delta)$ slopes by \Cref{prop:series-parallel}. By \Cref{le:nice}, we may further assume that $G$ is irreducible. 
We begin by proving the following.

\begin{lemma}\label{lem:twocut}
	Let $G$ be an irreducible $2$-connected cycle-tree. If the cycle of $G$ contains at least $4$ vertices, then any $2$-cut of $G$ consists of a tree-vertex and a cycle-vertex.
\end{lemma}
\begin{proof}
	First, we show that $G$ contains no $2$-cuts composed of pairs of tree-vertices. If such $2$-cut existed, removing its vertices would yield one component containing all the cycle-vertices and at least one component containing only tree-vertices. Such a component is either a path, which contradicts the fact that $G$ is irreducible, or it 
	contains a cut-vertex which would also be a cut-vertex in $G$, thus contradicting the fact that $G$ is $2$-connected.
	Second, we show that $G$ contains no $2$-cuts composed of pairs of cycle-vertices. If such $2$-cut existed, removing it would yield one component containing all the tree-vertices, and either at least two components containing cycle vertices, or exactly one component with at least two cycle vertices. Since the cycle of $G$ is chordless, both cases contradict the fact that $G$ is irreducible.
\qed\end{proof}

\begin{figure}[tbp]
	\centering
	\includegraphics[width=.35\columnwidth,page=1]{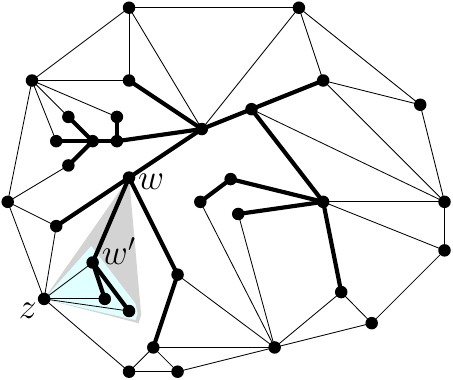}
	\caption{A $2$-connected cycle-tree with highlighted a $(w,z)$-flag (in gray) and a $(w',z)$-flag (in light blue). The 2-cut $\{w',z\}$ is dominated by the 2-cut $\{w,z\}$.}
	\label{fig:flags-1}
\end{figure}

Let $\{w,z\}$ be a $2$-cut of $G$, where $w$ is a tree-vertex and $z$ is a cycle-vertex. By removing $w$ and $z$ from $G$, we obtain $k \geq 2$ connected subgraphs $H_0, H_1, \dots, H_{k-1}$. The subgraph $C_i$ of $G$ induced by $V(H_i) \cup \{w, z\}$ is a component of $G$ with respect to $\{w,z\}$ ($0 \leq i \leq k-1$). One of such components, say $C_0$, contains all the cycle-vertices of $G$. The union of all components different from $C_0$ is called the \emph{$(w,z)$-flag} of $G$. See \cref{fig:flags-1} for an example.
Since $z$ has degree at most $\Delta$ and since $G$ is irreducible, we have the following.

\begin{property}\label{prop:constant-size-flags}
	For any $2$-cut $\{w,z\}$, the $(w,z)$-flag has $O(\Delta)$ vertices.
\end{property}

We say that a $2$-cut $\{w',z\}$, with $w' \neq w$, is \emph{dominated} by $\{w,z\}$ if $w'$ belongs to the $(w,z)$-flag of $G$. We say that  $\{w,z\}$ is \emph{dominant} when no other $2$-cut dominates it.
Let $G_2$ be the graph obtained from $G$ as follows: (i) remove, for each dominant $2$-cut $\{w,z\}$, all vertices of the $(w,z)$-flag of $G$ except $w$ and $z$, and add the edge~$(w,z)$, called the \emph{virtual edge} of $\{w,z\}$, if it does not already exist in $G$; (ii) contract all contractible vertices, if any. We call $G_2$ the \emph{$2$-frame graph} of $G$. Now we have removed all $(w,z)$-flags without introducing new ones. By \Cref{lem:twocut}, $G_2$ has no $2$-cuts and therefore it is a $3$-connected cycle-tree.

Let $e$ be an edge of a straight-line drawing $\Gamma$ and let $R$ be a rhombus whose longer diagonal is $e$. If $R \cap \Gamma=\{e\}$ we say that $R$ is a \emph{nice rhombus} for $e$.   

\begin{lemma}\label{lem:psn-biconnected-cycle-trees-graph}
	Every $2$-connected cycle-tree $G$ with maximum degree $\Delta$ has $\psn(G) \in O(\Delta^2)$.
\end{lemma}
\begin{proof}
	By \cref{le:nice} we can assume that $G$ is irreducible. We construct an embedding-preserving planar straight-line drawing $\Gamma$ of $G$ as follows.
	Let $G_2$ be the $2$-frame graph of $G$, and
	let $\Gamma_2$ be the planar straight-line drawing of $G_2$, obtained by applying \cref{lem:psn-triconnected-cycle-trees-graph} and by subdividing the edges that stemmed from the contraction operation (if any).	We define an angle $\beta > 0$ and for each virtual edge~$e=(w,z)$ of~$G_2$ we define a nice rhombus for $e$, such that the interior angles at $w$ and $z$ are both equal to $\beta$. Angle $\beta$ is chosen such that no two nice rhombi intersect each other (except at common corners). We then apply \cref{prop:series-parallel} to draw each $(w,z)$-flag inside the corresponding nice rhombus for $(w,z)$.
	Since no two nice rhombi intersect each other (except at common corners), the resulting drawing $\Gamma$ of $G$ is planar. Concerning the number of slopes, we have that $\Gamma_2$ uses
	$O(\Delta^2)$ slopes by \cref{lem:psn-triconnected-cycle-trees-graph}. We now argue that, overall, the $(w,z)$-flags use $O(\Delta^2)$ additional slopes. The drawing of each $(w,z)$-flag such that $\overline{wz}$ has slope $s$ in $\Gamma_2$ uses the~$O(\Delta)$ slopes in the set $\mathcal L(\beta,s,\Delta)$ of \Cref{prop:series-parallel}. Since each virtual edge of $G_2$ connects a tree-vertex and a cycle-vertex, by \cref{rem:red-slope-cycle} it never uses one of the red slopes. Hence the total number of slopes used by all virtual edges is $O(\Delta)$, which implies that, overall, the $(w,z)$-flags use $O(\Delta^2)$ slopes.
\qed\end{proof}

\subsection{$1$-Connected Cycle-Trees}\label{sec:one-connected}
In order to extend our construction to the $1$-connected case, we adopt a similar (but simpler) strategy as for the $2$-connected case.

\begin{figure}[tbp]
	\centering
	\includegraphics[width=.35\columnwidth,page=2]{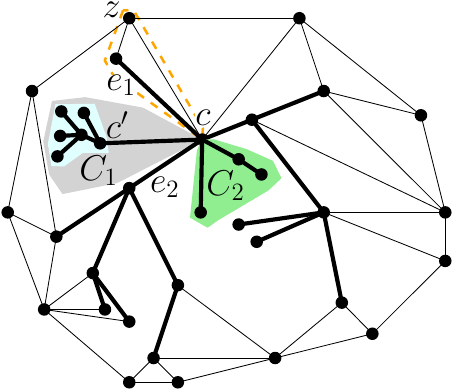}
	\caption{A $1$-connected cycle-tree with highlighted two distinct $c$-flags $C_1$ (in gray) and $C_2$ (in green) and a $c'$-flag (in light blue). The cut-vertex $c'$ is dominated by the cut-vertex $c$. The reference edge of $C_1$ is $e_1$ and the reference edge of $C_2$ is $e_2$. $C_1$ is a $c$-flag of  Type 1 because $e_1$ is an edge of a $(c,z)$-flag (highlighted with a dashed orange curve). $C_2$ is a $c$-flag of Type 2. }
	\label{fig:flags-2}
\end{figure}

Throughout this section $G$ is a $1$-connected cycle-tree.  
By \Cref{le:nice}, we may assume $G$ be irreducible.
Let $c$ be a cut-vertex of $G$.  By removing $c$ from $G$, we obtain $k \geq 2$ connected subgraphs $H_0, H_1, \dots, H_{k-1}$. The subgraph $C_i$ of $G$ induced by $V(H_i) \cup \{c\}$ is a component of $G$ with respect to $c$ ($0 \leq i \leq k-1$). One of such components, say $C_0$, contains all the cycle-vertices of $G$. Consider any pair of edges~$e_1$ and $e_2$ incident to $c$ that are consecutive in the counter-clockwise order around $c$ in $C_0$; the union of all components different from $C_0$ that have an edge incident to $c$ appearing between $e_1$ and $e_2$ in the counter-clockwise order of the edges around $c$ in $G$ is a \emph{$c$-flag of $G$}; $e_1$ is the \emph{reference edge} of the $c$-flag and $e_2$ is the \emph{second reference edge} of the $c$-flag. See \cref{fig:flags-2} for an example.
We say that a cut-vertex $c'$ is \emph{dominated} by $c$ if $c'$ belongs to the $c$-flag of $G$. A cut-vertex is \emph{dominant} when it is not dominated by any other cut-vertex.

Let $G_1$ be the graph obtained from $G$ as follows: (i) remove, for each dominant cut-vertex $c$, all vertices of the $c$-flags of $G$ except $c$; (ii) contract all contractible vertices, if any. We call $G_1$ the \emph{$1$-frame graph} of $G$. 
We have the following.

\begin{lemma}
	Let $G$ be an irreducible $1$-connected cycle-tree. The $1$-frame of $G$ is a $2$-connected cycle-tree.
\end{lemma}

\begin{proof}
	Let $G_1$ be the $1$-frame of $G$. Graph $G_1$ is $2$-connected since, by removing the $c$-flags, all the cut-vertices of $G$ are not cut-vertices of $G_1$; moreover no new cut-vertex has been introduced. Graph $G$ is also a cycle-tree since~$G$ is a cycle-tree and we only removed tree-vertices from $G$ that are not dominant cut-vertices.
\qed\end{proof}

The following lemma consider the special case when $G$ is a partial $2$-tree and it will be used in the proof of \Cref{thm:cycle-pseudotree}.

\begin{lemma}\label{le:g1}
	Let $G$ be an irreducible $1$-connected cycle-tree. If $G$ is a partial $2$-tree, its $1$-frame has $O(\Delta)$ edges.
\end{lemma}
\begin{proof}
	The $1$-frame graph $G_1$ of $G$ is a $2$-connected cycle-tree. Thus, removing the vertices of the outer boundary of $G_1$ one is left with a single tree $T$. Since $G$ is a partial $2$-tree,  $G_1$ is a ($2$-connected) series-parallel graph and therefore there exists exactly two vertices $u$ and $v$ of the outer boundary of $G_1$ that are adjacent to vertices of $T$. Since the outer boundary of $G_1$ is chordless, any other vertex of the outer boundary different from $u$ and $v$ has degree two. Since $G_1$ is irreducible there is only one such vertex. It follows that the outer boundary of $G_1$ is a $3$-cycle. Also, all tree-vertices of $G_1$ that have degree at most two in the tree are adjacent to $u$ or to $v$; since both $u$ and $v$ have degree at most $\Delta$ there are $O(\Delta)$ such vertices and hence the tree has $O(\Delta)$ vertices.
\qed\end{proof}

\begin{lemma}\label{le:1conn}
	Every $1$-connected cycle-tree $G$ with maximum degree $\Delta$ has $\psn(G) \in O(\Delta^2)$.
\end{lemma}
\begin{proof}
	By \cref{le:nice} we can assume that $G$ is irreducible. Furthermore, since removing the vertices of the outer boundary of $G$ must yield a tree, at most one cut-vertex of $G$ is a cycle-vertex. Also, if such a vertex exists, then $G$ is a partial $2$-tree and can be drawn with $O(\Delta)$ slopes by \Cref{prop:series-parallel}. Hence we shall assume that every cut-vertex is a tree-vertex.
	
	We construct a planar straight-line drawing $\Gamma$ of $G$ as follows.
	Let $G_1$ be the $1$-frame graph of $G$, and
	let $\Gamma_1$ be the planar straight-line drawing of $G_1$, obtained by applying \cref{lem:psn-biconnected-cycle-trees-graph} and by subdividing the edges that stemmed from the contraction operation (if any).
	
	In the following, we assume that $G \neq G_1$, as otherwise $G$ is $2$-connected, and simply setting $\Gamma = \Gamma_1$ proves the statement. 
	We now show how to insert the $c$-flags into $\Gamma_1$ so to construct $\Gamma$. We distinguish  between the $c$-flag whose reference edge belongs to some $(w,z)$-flag, which we call \emph{$c$-flags of Type 1}, and those whose reference edge belongs to the $2$-frame $G_2$, which we call \emph{$c$-flags of Type 2} (see \cref{fig:flags-2}). Let $T_c$ be any $c$-flag of Type 1 and let $G_{w,z}$ be the $(w,z)$-flag that contains the reference edge of $T_c$. Observe that $T_c \cup G_{w,z}$ is a partial $2$-tree. Let~$R$ be the nice rhombus for $(w,z)$ defined in the proof of \Cref{lem:psn-biconnected-cycle-trees-graph}. We delete from~$\Gamma_1$ the drawing of $G_{w,z}$ and apply \Cref{prop:series-parallel} to draw $T_c \cup G_{w,z}$ inside $R$. Let $\Gamma'_1$ be the drawing obtained once all the Type 1 $c$-flags have been processed. 
	
	We now add to $\Gamma'_1$ the Type 2 $c$-flags. For every $c$-flag we suitably identify an edge $e_c$ as follows. Let $T_c$ be a Type 2 $c$-flag, let $e_1$ be the reference edge of $T_c$, and let $e_2$ be the second reference edge of $\Gamma'_1$. If the slope of $e_1$ is non-red, then $e_c=e_2$; if the slope of $e_1$ is red and the slope of $e_2$ is non-red then $e_c=e_2$; otherwise, $e_c$ is any edge of $T_c$ incident to $c$. 
    Notice that, in the latter case, $e_c$ is not an edge of $\Gamma'_1$ and $e_1$ and $e_2$ are drawn with two red slopes. Let $W$ be the wedge swept by rotating $e_1$ counterclockwise until it overlaps $e_2$. By \Cref{remark:non-consecutive-red-slope} there exists a non-red slope $s$ in the set $\mathcal S$ such that the ray $r$ originating at $c$ having slope $s$ lies inside $W$. We draw edge $e_c$ in $\Gamma'_1$ along ray $r$ such that it does not intersect any other edges.  We define an angle $\beta > 0$ and for each edge $e_c=(c,w)$ of every Type~2 $c$-flag $T_c$ we define a nice rhombus for $e_c$, such that the interior angles at $c$ and $w$ are both equal to $\beta$. Angle $\beta$ is chosen such that no two nice rhombi intersect each other. Since $T_c \cup e_c$ is a tree (and hence a partial $2$-tree), we can apply \cref{prop:series-parallel} to draw the Type 2 $c$-flag inside the nice rhombus for $e_c$.

	Since no two nice rhombi intersect each other (except at common corners), the resulting drawing $\Gamma$ of $G$ is planar. Concerning the number of slopes, we have that~$\Gamma_1$ uses
	$O(\Delta^2)$ slopes by \cref{lem:psn-biconnected-cycle-trees-graph}. We now argue that, overall, the $c$-flags use $O(\Delta^2)$ additional slopes. All the nice rhombi used to draw the Type 1 and Type~2 $c$-flags are defined for edges that have a non-red slope $s \in \mathcal S$, that is for edges with $O(\Delta)$ different slopes in total. For each such nice rhombus, the drawing of the $c$-flag inside the rhombus uses the $O(\Delta)$ slopes in the set $\mathcal L(\beta,s,\Delta)$ of \Cref{prop:series-parallel}. Hence, the drawings of all $c$-flags use $O(\Delta^2)$ slopes overall.
 \qed\end{proof}

\Cref{le:1conn,le:nice} imply the following.

\begin{theorem}\label{lem:one-connected-construction}
	Every cycle-tree $G$ with maximum degree $\Delta$ has $\psn(G) \in O(\Delta^2)$.
\end{theorem}

\section{Nested Pseudotrees}\label{se:nested}

\begin{figure}[htbp]
	\centering
	\includegraphics[width=.35\columnwidth,page=3]{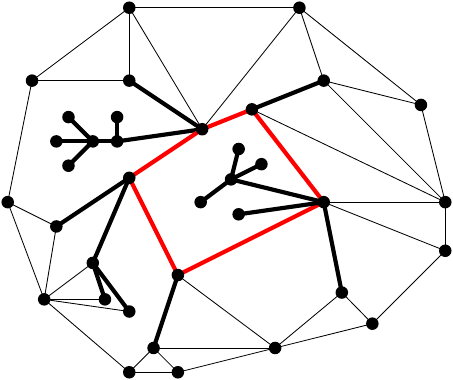}
	\caption{A cycle-pseudotree. The edges of the pseudotree are bold and the cycle of the pseudotree is red.}
	\label{fig:cycle-pseudotree}
\end{figure}

To prove \cref{thm:main}, we first consider nested-pseudotrees whose outer boundary is a chordless cycle. We call such graphs \emph{cycle-pseudotrees} (see~\cref{fig:cycle-pseudotree} for an example). 

\subsection{Cycle-Pseudotrees}\label{apx:cycle-pseudotree}

Let $H$ be a degree-$\Delta$ cycle-pseudotree graph with pseudotree $P$. Every edge of the unique cycle of $P$ is called a \emph{disposable edge} of $H$.
Let $G$ be the graph obtained by removing a disposable edge $e=(u,v)$ from $H$. Clearly, $G$ is a cycle-tree, and~$u$ and~$v$ are tree-vertices of $G$. 
Let $C$ be a $c$-flag of a cut-vertex $c$ of $G$ and let~$x$ be a tree-vertex of $C$ different from $c$; we say that $x$ \emph{belongs to a $c$-flag} of $G$. Analogously, let $W$ be the $(w,z)$-flag of a $2$-cut $\{w,z\}$ of $G$ and let $x$ be a tree-vertex of $W$ different from $w$ (recall that $z$ is a cycle vertex); we say that $x$  \emph{belongs to the $(w,z)$-flag} of~$G$.

\begin{theorem}\label{thm:cycle-pseudotree}
	Every cycle-pseudotree $H$ with maximum degree $\Delta$ has $\psn(H) \in O(\Delta^2)$.
\end{theorem}
\begin{proof}
	Let $P$ be the pseudotree of $H$ and et $e=(u,v)$ be any disposable edge of $H$, let $G=H \setminus \{e\}$, let $G_1$ be the $1$-frame of $G$, and let  $G_2$ be the $2$-frame of $G$. 
	We distinguish cases based on the endpoints of $e$.

	\smallskip
	\noindent {\bf Case A.} There exists a dominant cut-vertex $c$ of $G$ such that $u$ belongs to a $c$-flag $C$. Observe that since $H$ is a plane graph, $v$ cannot belong to some $c'$-flag $C'$ distinct from $C$ and with $c'=c$. Hence, we distinguish the following subcases: 
	
	\begin{inparaenum}[\bf {A.}1]
	
		\item  $v$ belongs to the $c$-flag $C$,
	
		\item  $v$ belongs to a $c'$-flag $C'$ of some dominant cut-vertex $c'\neq c$,
	
		\item  $v$ belongs to the $(w,z)$-flag of a dominant $2$-cut $\{w,z\}$ of $G_1$, 
			
		\item  $v$ belongs to $G_2$.
	
	\end{inparaenum}
	
	\smallskip
	If Case A does not apply, then we may assume that neither $u$ nor $v$ belongs to a $c$-flag of any cut-vertex $c$ of $G$. That is, both $u$ and $v$ belong to $G_1$. 
	
	\smallskip
	\noindent {\bf Case B.} There exists a dominant $2$-cut $\{w,z\}$  of $G_1$ such that $u$ belongs to the~$(w,z)$-flag. Let $W$ denote this $(w,z)$-flag.  We distinguish three subcases: 
	
	\begin{inparaenum}[\bf {B.}1]
		
		\item  $v$ belongs to the $(w,z)$-flag $W$,
		
		\item  $v$ belongs to the $(w',z')$-flag $W'$ of some dominant $2$-cut $\{w',z'\} \neq \{w,z\}$ of $G_1$,
		
		\item  $v$ belongs to $G_2$.
	
	\end{inparaenum}
	
	\smallskip
	\noindent {\bf Case C.} If Case~A and Case~B do not apply, then both $u$ and $v$ belong to $G_2$.

	We now show how to obtain a planar straight-line drawing $\Gamma$ of $H$ using $O(\Delta^2)$ slopes in each of the above cases. We obtain $\Gamma$ recursively. Each of the cases yields either a smaller instance to which a different case applies or it is a base case (i.e., {\bf A.1}, {\bf B.1}, and {\bf C}) in which we use \cref{lem:one-connected-construction} to obtain a planar straight-line drawing using $O(\Delta^2)$ slopes. Crucially in all cases the depth of the recursion is constant and each recursive call increases the number of slopes by $O(\Delta)$. 
	
	In {\bf Case~A.1}, both $u$ and $v$ belong to the $c$-flag $C$ of some cut-vertex $c$ of $G$. Refer to \cref{fig:cases-pseudotree-a.1}. We have that $C$ together with the edge $(u,v)$ forms a pseudotree, and thus a partial $2$-tree. Hence, $C$ can be drawn exploiting \cref{prop:series-parallel}, and thus~$\Gamma$ can be obtained by applying the algorithm in the proof of \cref{le:1conn} without any modification (after contracting all contractible vertices, if any).
	
	\begin{figure}[t!]
		\centering
		\includegraphics[width=\textwidth,page=28]{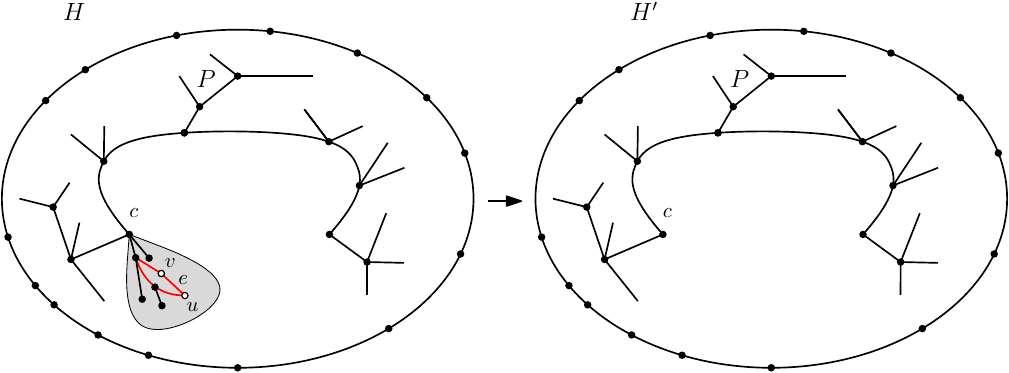}
		\caption{Illustration for Case A.1 of
			\cref{thm:cycle-pseudotree}. The cycle of $P$ is red. The $c$-flag is highlighted with a grey background.}
		\label{fig:cases-pseudotree-a.1}
	\end{figure}
	
	In {\bf Case~A.2}, {\bf A.3}, and {\bf A.4}, $u$ belongs to $C$ and $v$ does not. Observe that, $c$ and $v$ are not cycle-vertices. Refer to \cref{fig:cases-pseudotree-a}. Let $H'$ be the graph obtained by removing from $G$ the vertices belonging to $C$ and by inserting the edge $(c,v)$, if it is not in $G$ already. Note that, $H'$ is a cycle-pseudotree. In fact, $c$ and $v$ are vertices of the cycle of $P$. We obtain a drawing $\Gamma'$ of $H'$ as follows. If $H'$ is a cycle-tree we draw it by applying \Cref{lem:one-connected-construction}; otherwise $H'$ is a cycle-pseudotree containing less vertices than $H$, and thus the drawing $\Gamma'$ of $H'$ can be obtained recursively using $(c,v)$ as the disposable edge. We now modify $\Gamma'$ to obtain $\Gamma$.  Let $H_{e}$ be the union of the $c$-flag $C$, the vertex $v$, and the edges $(c,v)$ and $(u,v)$. Clearly,  $H_{e}$ is a pseudotree, and thus a partial $2$-tree. Let $R$ be a nice rhombus for $(c,v)$ in $\Gamma'$. We draw $H_{e}$ inside $R$ by applying \cref{prop:series-parallel} using $O(\Delta)$ additional slopes. Finally, we remove the edge $(c,v)$, if it is not in $H$. This provides $\Gamma$.
	
	We show that the recursion moves to Case~B or to Case~C in at most two steps. If {\bf Case~A.2} applies, then $v$ still belongs to the $c'$-flag $C'$ in $H' \setminus (c,v)$ and $c$ does not belong to any $x$-flag of a cut-vertex $x$ of $H' \setminus (c,v)$. Therefore, if $c$ belongs to the $(w,z)$-flag of some $2$-cut $\{w,z\}$ of $H' \setminus (c,v)$, and thus of $G_1$, then we recurse to {\bf Case~A.3}, otherwise $c$ belongs to $G_2$, and we recurse to {\bf Case~A.4}. If {\bf Case~A.3} applies, then we recurse to one of {\bf Case~B.1}, {\bf B.2}, and {\bf B.3}. If {\bf Case~A.4} applies, then we recurse to one of {\bf Case B.3} and {\bf Case C}.
	
	\begin{figure}[t!]
		\centering
		\includegraphics[width=\textwidth,page=24]{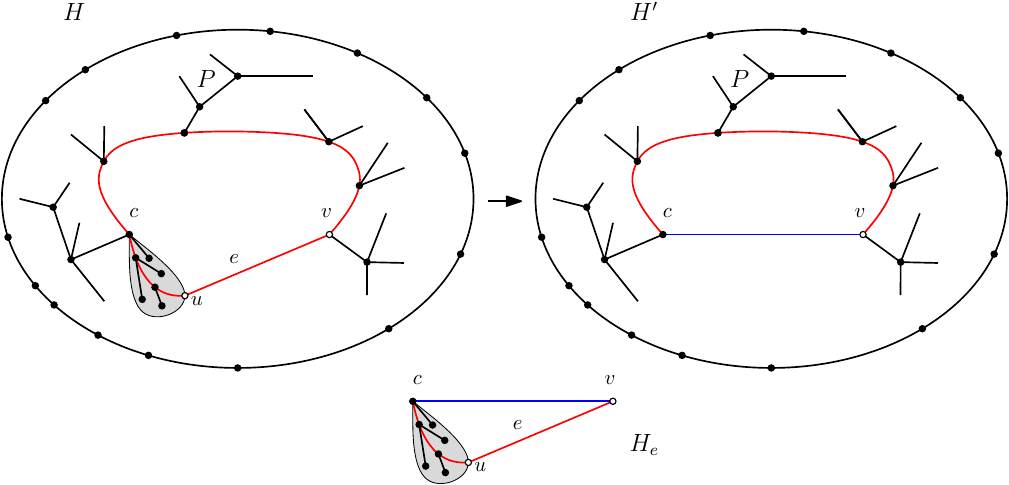}
		\caption{Illustration for Cases A.2, A.3, and A.4 of
			\cref{thm:cycle-pseudotree}. The cycle of $P$ is red. The $c$-flag is highlighted with a grey background.}
		\label{fig:cases-pseudotree-a}
	\end{figure}
	
	In {\bf Case~B}, both $u$ and $v$ are vertices of $G_1$. We will construct a drawing $\Gamma_1$ of the graph $G_1 \cup (u,v)$. Then, $\Gamma$ is obtained from  $\Gamma_1$ by drawing all the $c$-flags of $G$ using $O(\Delta)$ new slopes, as described in the proof of \cref{le:1conn}.

	\begin{figure}[t]
		\centering
		\includegraphics[width=\textwidth,page=29]{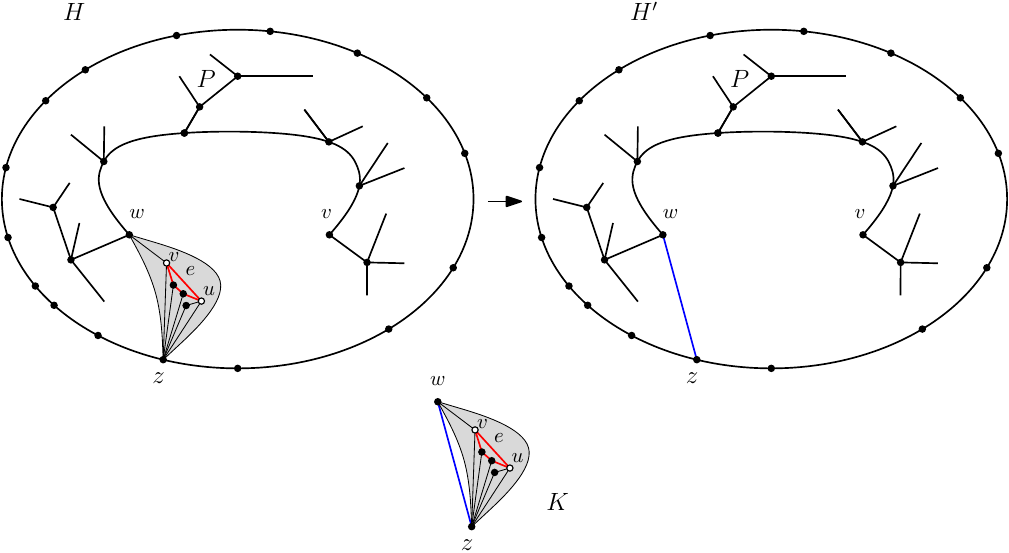}
		\caption{Illustration for Case B.1 of
			\cref{thm:cycle-pseudotree}. The cycle of $P$ is red. The $(w,z)$-flag is highlighted with a grey background.}
		\label{fig:cases-pseudotree-b.1}
	\end{figure}

	In {\bf Case~B.1}, both $u$ and $v$ belong to the $(w,z)$-flag $W$. Refer to \cref{fig:cases-pseudotree-b.1}.  We have that the $W$ together with the edge $(u,v)$ forms a planar graph $K$ containing~$O(\Delta)$ vertices, by \cref{prop:constant-size-flags}. Let $H'$ be the graph obtained by removing the vertices belonging to $W$ from $G$, and by inserting the edge $(w,z)$, if it is not in $G$ already. Note that, $H'$ is a cycle-tree, because the cycle of $P$ belongs to $K$. First, we construct a drawing $\Gamma'$ of $H'$ by applying~\cref{lem:one-connected-construction}.
	Then, $\Gamma_1$ can be obtained from $\Gamma'$ by drawing $K$ inside a nice rhombus for the edge $(w,z)$ in $\Gamma'$ by using the classical Tutte's algorithm~\cite{Tutte}, and by removing the edge $(w,z)$, if it is not in $G$. This can be done with $O(\Delta)$ additional slopes because $K$ has size $O(\Delta)$.
	
	In {\bf Case~B.2} and {\bf B.3}, $u$ belongs to the $(w,z)$-flag $W$ of $G_1$ and $v$ does not. Refer to \cref{fig:cases-pseudotree-b}. Let $H'$ be the graph obtained by removing from $G$ the vertices belonging to $W$ and to the $(v,z)$-flag (if it exists), and by inserting the edges~$(w,v)$, $(z,v)$, and $(w,z)$, if they do not already belong to $G$.  Note that, $H'$ is a cycle-pseudotree. In fact, $w$ and $v$ are vertices of the cycle of $P$. We obtain a drawing $\Gamma'$ of $H'$ as follows. If $H'$ is a cycle-tree we draw it applying \Cref{lem:one-connected-construction}; otherwise~$H'$ is a cycle-pseudotree containing less vertices than $H$, and thus the drawing $\Gamma'$ of $H'$ can be obtained recursively using $(w,v)$ as the disposable edge.  We now modify $\Gamma'$ to obtain $\Gamma_1$ as follows. Let $H_{e}$ be the union of $W$, the vertex $v$, and the edges~$(w,v)$,~$(z,v)$, $(w,z)$, and $(u,v)$. Notice that, by \cref{prop:constant-size-flags}, $H_e$ has size~$O(\Delta)$. Then, $\Gamma_1$ is obtained from $\Gamma'$ as follows: $H_e$ is drawn inside the triangle~$(w,v)$, $(z,v)$, and $(w,z)$ by Tutte's algorithm~\cite{Tutte}; if the $(v,z)$-flag exits it is drawn by \Cref{prop:series-parallel} inside a nice rhombus for $(v,z)$; finally edges $(w,v)$, $(z,v)$, and $(w,z)$ are removed, if they are not in $G$. 
	
	\begin{figure}[t]
		\centering
		\includegraphics[width=\textwidth,page=25]{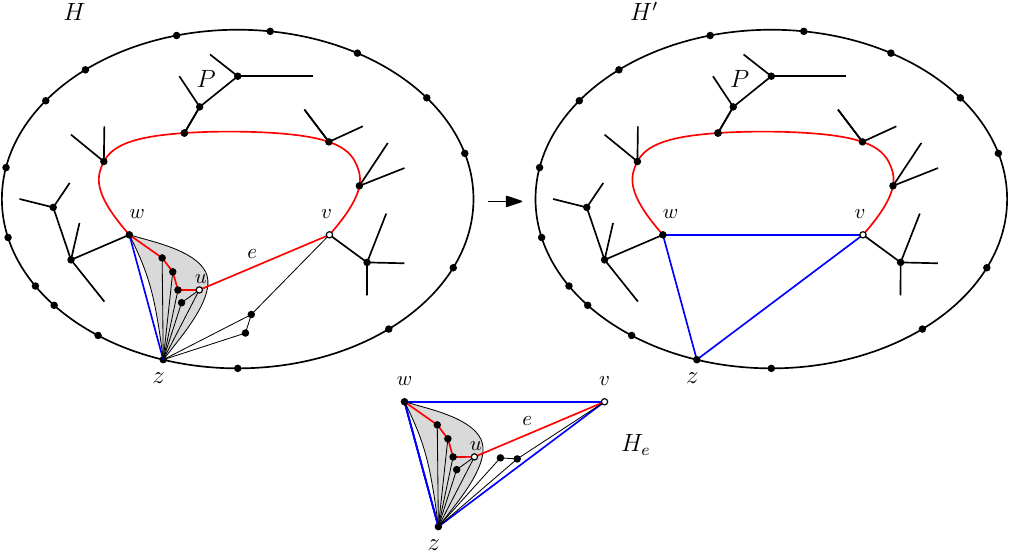}
		\caption{Illustrations for Cases B.2 and B.3 of
			\cref{thm:cycle-pseudotree}. The cycle of $P$ is red. The $(w,z)$-flag is highlighted with a grey background.}
		\label{fig:cases-pseudotree-b}
	\end{figure}

	We show that the recursion moves to Case~C in at most two steps. 
	If {\bf Case~B.2} applies, then $v$ still belongs to the $(w',z')$-flag $W'$ in $H' \setminus (w,v)$ and $w$ belongs to the $2$-frame of $H' \setminus (w,v)$. Therefore, we recurse to {\bf Case~B.3}. 
	If {\bf Case~B.3} applies, then both $w$ and $v$ belong to the $2$-frame of $H' \setminus (w,v)$, and we recurse to {\bf Case~C}.

	In {\bf Case~C}, both $u$ and $v$ are in $G_2$. In order to construct $\Gamma$ we proceed as follows. We first construct a drawing $\Gamma_2$ of the graph $G_2 \cup (u,v)$. Then, we obtain a drawing $\Gamma_1$ of $G_1 \cup (u,v)$ by adding to $\Gamma_2$  all the $(w,z)$-flags of $G$ as described in the proof of \cref{le:1conn}, which uses $O(\Delta^2)$ new slopes. Finally, we obtain $\Gamma$ from~$\Gamma_1$ by adding all the $c$-flags of $G$ as described in the proof of \cref{le:1conn}, which uses~$O(\Delta)$ new slopes. From this discussion it suffices to show how to compute $\Gamma_2$. 
	
	\begin{figure}[h!]
		\centering
		\subfigure[]{\includegraphics[	width=\textwidth,page=1]{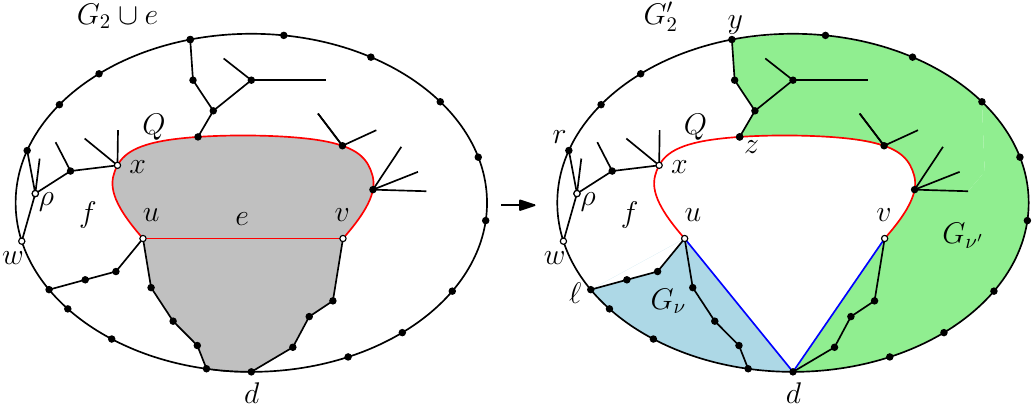}\label{fig:cases-pseudotree}}
		\subfigure[]{\includegraphics[	width=0.6\textwidth,page=2]{figs/final-figure.pdf}\label{fig:cases-pseudotree-2}}
		\caption{Illustrations for Case C of \cref{thm:cycle-pseudotree}. (a) Construction of $G_2'$ from $G_2 \cup e$; in the left part, the shaded region is face $g$. The cycle of $P$ is red. (b) The path-tree $G^*$ rooted at $\rho$}
		\label{fig:case-C-pseudotree}
	\end{figure}
	
	Refer to \cref{fig:case-C-pseudotree}.  Let $g$ be the unique face of $G_2$ that is incident to $u$ and $v$ (this face is unique because otherwise $u$ and $v$ would be a $2$-cut, but this is not possible by \Cref{lem:twocut}). Let $d$ be any cycle-vertex of $g$. Let $G'_2$ be the graph obtained by adding the edges $(u,d)$ and $(v,d)$ to $G_2$. Observe that, $G'_2$ is a cycle-tree and, like $G_2$, it is $3$-connected.
	Let $Q$ be the path between $u$ and $v$ in the tree of $G'_2$.
	Let $x$ be the neighbor of $u$ in $Q$. Consider the face $f$ of $G'_2$ having the edge $(u,x)$ on its boundary that does not contain $v$. Let $w$ be the first cycle-vertex that is encountered when traversing the boundary of $f$ starting from $x$ and avoiding $u$. Let $\rho$ be the tree-vertex preceding $w$ in such a traversal (notice that $\rho$ may coincide with $x$). 
	Let $G^*$ be the path-tree illustrated in~\cref{fig:case-C-pseudotree}. $G^*$ is constructed from $G'_2$ by removing $w$ from the cycle-tree and by choosing $\rho$ as the root vertex of $G^*$ (see also the construction before the proof \cref{lem:psn-triconnected-cycle-trees-graph}). 
	Let $\Gamma^*$ be the planar straight-line drawing of $G^*$ inside an equilateral triangle, obtained by applying \cref{lem:psn-almost-$3$-connected}.
	Let $\Gamma'_2$ be a planar straight-line drawing of $G'_2$ such that
	$\psn(\Gamma'_2) \leq \psn(\Gamma^*) + \Delta$, obtained by applying \cref{lem:psn-triconnected-cycle-trees-graph} starting from $\Gamma^*$.
	By our selection of $\rho$, we have that in any SPQ-tree of $G^*$ the vertex $x$ is the root of a P-node $\mu$ having two consecutive children $\nu$ and $\nu'$ sharing the cycle-vertex~$d$. Moreover, we have that the edge $(u,d)$ belongs to the right path $R$ of the pertinent graph $G_{\nu}$ of $\nu$, and that the edge $(v,d)$ belongs to the left path $L$ of the pertinent graph $G_{\nu'}$ of $\nu'$, unless they have been contracted because they had degree $2$ after the removal of~$(u,v)$. In the latter case, if $u$ (resp. $v$) has been contracted we first reinsert it in the drawing by subdividing the edge incident to $d$ that belongs to $R$ (resp. to~$L$).
	Since, in $\Gamma^*$, the path $R$ is $\leftx$-monotone and the path $L$ is $\rightx$-monotone, except possibly for the last edges of such paths incident to the root $x$  of $\mu$, it is possible to draw the edge $(u,v)$ in $\Gamma^*$, and thus in $\Gamma'_2$, without introducing any crossings, possibly using an additional slope. This concludes the construction of the drawing~$\Gamma_2$ of $G_2 \cup (u,v)$. Drawings $\Gamma_1$ and $\Gamma$ can, in fact, be obtained starting from $\Gamma_2$ as previously described.
\qed\end{proof}

\subsection{Proof of \cref{thm:main}}

Let $G$ be a nested pseudotree of degree $\Delta$. If $G$ is a cycle-pseudotree, we are done by \Cref{thm:cycle-pseudotree}. Thus, assume otherwise.
By definition, removing the vertices on the outer face of $G$ yields a pseudotree $P$. 
Let $C$ be the chordless cycle of $G$ that contains $P$ in its interior. Denote the vertices of $C$ by $u_0,u_1,\dots,u_{|C|-1},u_{|C|}=u_0$ in the order in which they appear in a clockwise visit of $C$.
If we remove $C$ from~$G$, then $G$ is decomposed into components $G_0, G_1, \dots, G_h$, such that one of them, say $G_0$, coincides with $P$, while every other component is an outerplanar graph. For $i=1,\dots,h$, each $G_i$ is connected to $C$ by edges that are incident to either a common vertex $u_k$ or to a common pair $u_k,u_{k+1}$ of adjacent vertices of $C$, for some~$0 \le k \le |C|-1$. In both cases, we refer to $(u_k,u_{k+1})$ as the \emph{base edge} of~$G_i$. Note that, each $G_i$ has a unique base edge, but different $G_i$'s may share the same base edge.
For $k=0,\dots, |C|-1$, let $G^+_k$ denote the subgraph of $G$ induced by the union of $\{u_{k}, u_{k+1}\}$ and of the vertex sets of the graphs $G_i$ whose base edge is $(u_k,u_{k+1})$.
Note that each $G_k^+$ is an outerplane graph that contains the edge~$(u_k, u_{k+1})$ in its outer face. Let $G^*$ be the cycle-pseudotree defined as the subgraph of $G$ induced by the union of $C$ and $P$. We compute a planar straight-line drawing $\Gamma^*$ of $G^*$ using $O(\Delta^2)$ slopes by using \Cref{thm:cycle-pseudotree}. We can define a set of $|C|$ similar nice  triangles, one for each base edge and use \Cref{prop:series-parallel} to draw $G_k^+$ inside the corresponding triangle. The slope of all base edges, except two, is black. Hence every $G_k^+$ can be drawn by using the same set of $O(\Delta)$ slopes, except for two which require a rotation of the slopes. It follows that the planar slope number of $G$ is $O(\Delta^2)$.

\section{Halin Graphs}\label{se:halin}

We observe that  Halin graphs are $3$-connected cycle-trees with $\delta^*{=}3$ because each path-vertex has two incident edges that are incident to the outer face and it is a leaf when these two edges are removed. By Equation~\ref{eq:numberOfslopes}, we obtain~$|{\cal S}|{=}12\Delta{+}10$. Thus, \Cref{lem:psn-triconnected-cycle-trees-graph} implies that Halin graphs have planar slope number $\Theta(\Delta)$. 

\medskip

\begin{figure}[htbp]
	\centering
	\subfigure[]{\label{fig:slope-set2.a}\includegraphics[width=0.19\textwidth,page=1]{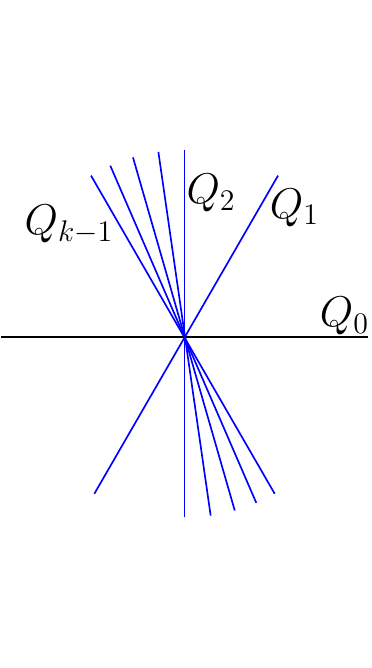}
	}
    \hfil
	\subfigure[]{\label{fig:slope-set2.b}\includegraphics[width=0.38\textwidth,page=3]{figs/slope-set2.pdf}
	}
	\caption{\label{fig:slope-set2.1}(a) A set of slope $\mathcal{S}_k$ for $k=7$; (b) Definition of the equilateral triangles for the recursive construction of a tree drawing that uses the slope set $\mathcal{S}_k$ of (a).}
\end{figure}

We now prove a finer upper bound for Halin graphs, namely, we show that Halin graphs have planar slope number at most $\Delta$ for $\Delta \geq 4$ and at most $4$ if $\Delta=3$. To this aim, we define a set of $k=\max\{4,\Delta\}$ slopes $\mathcal S_k$ as follows: $\mathcal S_k$ contains the slope 0, the slope $\frac{\pi}{3}$, the slope $\frac{\pi}{2}$, and the slope $\frac{2\pi}{3}$. If $\Delta>4$, we need $\Delta-4$ additional slopes. While our construction works for any set of $\Delta-4$ additional slopes whose value is between $\frac{\pi}{3}$ and slope $\frac{2\pi}{3}$, to simplify the description we  arbitrarily choose these $\Delta-4$ additional slopes between $\frac{\pi}{2}$ and $\frac{2\pi}{3}$ (see \Cref{fig:slope-set2.a}). Let $Q_0, Q_1, \dots, Q_{k-1}$ denote the slopes of $\mathcal S_k$ in increasing value. Notice that, by our choice of the slopes, $Q_0$ is the slope $0$ and $Q_2$ is the slope $\frac{\pi}{2}$. We will exploit the following technical lemma.

\begin{lemma}\label{lem:tree-in-triangle}
    Let $T$ be an $n$-vertex rooted ordered tree such that each vertex of $T$ has at least $2$ and at most $d$ children,  $d \geq 2$. Let $\rho$ be the root of $T$ and, if $n>1$, let $\ell$, and $r$  the leftmost leaf, and the rightmost leaf of $T$, respectively. Let $\bigblacktriangle(abc)$ be an equilateral triangle such that the segment $\overline{bc}$ is horizontal, $a$ lies above $\overline{bc}$, and $a$, $b$, and $c$ appear in this counter-clockwise order. Tree $T$ admits a straight-line order-preserving planar drawing $\Gamma$ such that: (i) $\Gamma$ uses the slopes in the set $\mathcal S_k$, where $k=\max\{4,d+1\}$; (ii) $\Gamma$ is contained in  $\bigblacktriangle(abc)$; (iii) if $n>1$, then $\rho$, $\ell$, and $r$ are mapped to the points $a$, $b$, and $c$ in $\Gamma$, respectively; (iv) if $n=1$, then $\rho$ is represented by a point at the intersection of $\overline{bc}$ with a straight line passing through $a$ and having any slope $Q_i\in \mathcal{S}_k \setminus \{S_0\}$; (v)  all the leaves of $T$ lie on $\overline{bc}$ in $\Gamma$.
\end{lemma}
\begin{proof}
The proof is by induction on the number of vertices $n$ of $T$. If $n=1$, then we can choose one of the slope $Q_i$ with $i>0$ and place the unique vertex of $T$ at the intersection point between $\overline{bc}$ and a straight line through $a$ with slope $Q_i$. Properties (i), (ii), (iv), and (v) hold by construction, while (iii) does not apply.

If $n>1$, let $T_1, T_2, \dots, T_g$, the (at most $d$) sub-trees of $T$ rooted at the children of $\rho$ in their left-to-right order. Let $\delta(b,c)$ be the length of $\overline{bc}$ and let $x$ be a value smaller than $\frac{\delta(b,c)}{2d}$. 
For every $i=1,2,\dots,d$ we define an equilateral triangle $\bigblacktriangle(a_ib_ic_i)$ with sides of length $x$ and such that: (i) the point $a_i$ is a point of the straight line $l_i$ having slope $Q_i$ and passing through the point $a$; (ii) the segment $\overline{b_ic_i}$ is contained in the segment $\overline{bc}$. 
By the choice of $x$, every triangle $\bigblacktriangle(a_ib_ic_i)$ is contained in the triangle $\bigblacktriangle(abc)$ and $\bigblacktriangle(a_ib_ic_i) \cap \bigblacktriangle(a_jb_jc_j)=\emptyset$, for $1 \leq i \neq j \leq d$. See \Cref{fig:slope-set2.b} for an illustration.

To construct the drawing $\Gamma$ of $T$ we recursively compute a drawing $\Gamma_i$ of each sub-tree $T_i$ inside the triangle $\bigblacktriangle(a_ib_ic_i)$, for $i=1,2,\dots,g-1$, and a drawing $\Gamma_g$ of $T_g$ inside the triangle $\bigblacktriangle(a_db_dc_d)$. We then place the root $\rho$ of $T$ at point $a$ and connect it to the points $a_1,a_2,\dots,a_{g-1}$ and $a_d$. Let $T_j$ be the sub-tree drawn inside $\bigblacktriangle(a_ib_ic_i)$ ($i$ is equal to $j$ if $j\leq g-1$, while $i$ is equal to $d$ if $j=g$). By induction, the point $a_i$ represents the root $\rho_j$ of $T_j$ if $T_j$ has more than one vertex. This means that, in this case, the edge $(\rho,\rho_j)$ is drawn as a segment using the slope $Q_i$ in $\mathcal S_k$. If $T_j$ has only one vertex, then, by property (iv), the single vertex $\rho_j$ of $T_j$ can be represented by a point $q$ at the intersection of the segment $\overline{b_ic_i}$ with the straight line passing through $a_i$ and having slope $Q_i$. Thus, also in this case the edge $(\rho,\rho_j)$ is represented by a segment with slope $Q_i$. It follows that all edges incident to $\rho$ are drawn with slopes in the set $\mathcal{S}_k$. Since by induction the edges of each $\Gamma_i$ use slopes in the set $\mathcal{S}_k$ property (i) holds for $\Gamma$. Property (ii) holds because each $\Gamma_i$ is  contained inside its assigned triangle and each such triangle is contained inside $\bigblacktriangle(abc)$. About property (iii) observe that, the leftmost leaf $\ell$ of $T$ coincides with the leftmost leaf of $T_1$ if $T_1$ has more than one vertex, otherwise it coincides with the single vertex of $T_1$. In the former case such a vertex is represented by the point $b_1$, which coincides with $b$; in the latter case, the unique vertex of $T_1$ is represented by a point $q$ that is the intersection of $\overline{b_1c_1}$ with a straight line passing through $a_1$ with slope $Q_1$; since the left side of $\bigblacktriangle(a_1b_1c_1)$ has slope $Q_1$ and by construction $a_1$ belongs to segment $\overline{ab}$, point $q$ coincides with $b$, and the leftmost leaf of $T$ is represented by $b$ also in this case. Analogously, $r$ either coincides with the rightmost leaf of $T_g$ or with the single vertex of $T_g$. With a symmetric argument as the one used for $\ell$, we can show that in both cases $r$ is represented by the point $c_d$, which coincides with $c$. Property (iv) does not apply in this case. Property (v) holds by induction and by the fact that each segment $\overline{b_ic_i}$ is contained in the segment $\overline{bc}$. 
\qed\end{proof}

\begin{figure}[htbp]
	\centering
	\subfigure[]{\label{fig:slope-set2.c}\includegraphics[width=0.38\textwidth,page=4]{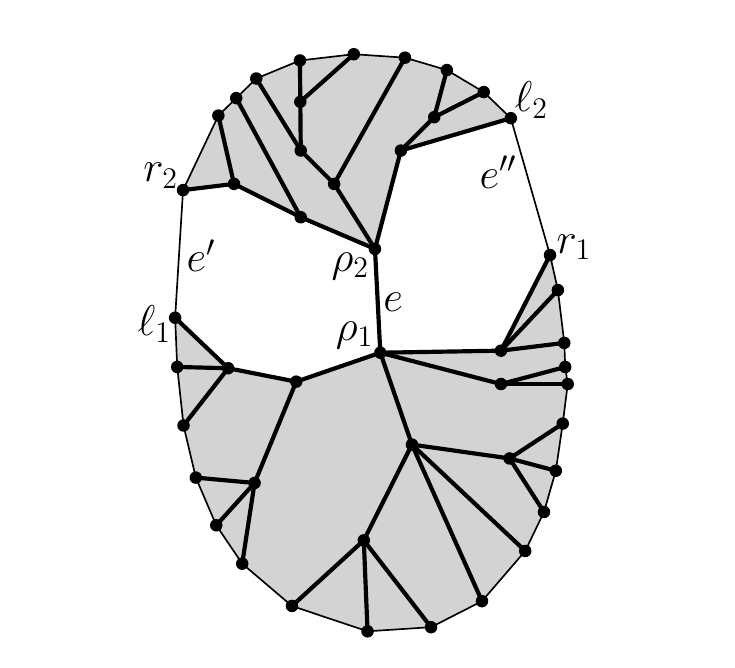}
	}
	\hfil
	\subfigure[]{\label{fig:slope-set2.d}\includegraphics[width=0.38\textwidth,page=5]{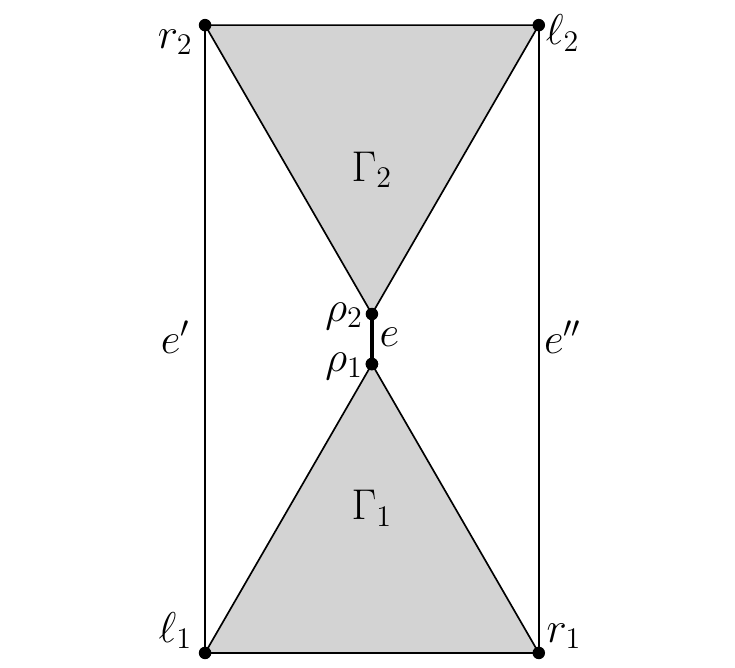}
	}
	\caption{\label{fig:slope-set2.2}(a) Decomposition of a Halin graph that has at least two internal vertices;  (b) Construction of a drawing of a Halin graph by combining two sub-drawings $\Gamma_1$ and $\Gamma_2$.}
\end{figure}

\subsubsection{Proof of \Cref{thm:main-halin}}. Let $G$ be a Halin graph different from $K_4$. We distinguish two cases depending on the number of internal vertices of $G$. If $G$ has only one internal vertex, i.e., it is a wheel with $n-1$ external vertices, we compute a drawing $\Gamma_n$ as follows. If $n=5$, the drawing $\Gamma_n$ is obtained by placing the four external vertices at the four corners of a square and the single internal vertex at the center of the square. The number of slopes of $\Gamma_5$ is clearly $4$. If $n>5$, then $\Gamma_n$ is obtained from $\Gamma_{n-1}$ by adding a vertex in any point of the outer quadrangle of $\Gamma_{n-1}$ and connecting it to the center of the wheel. Since $\Gamma_5$ uses $4$ slopes and each $\Gamma_n$ uses one slope more than $\Gamma_{n-1}$, the number of slopes of $\Gamma_n$ is $n-1$, which is equal to~$\Delta$. 

Assume now that $G$ has at least two internal vertices and therefore at least one edge $e=(\rho_1, \rho_2)$ such that both $\rho_1$ and $\rho_2$ are internal. The edge $e$ is incident to two faces each one having a single edge incident to the outer face of $G$. Let $e'=(\ell_1,r_2)$ and $e''=(\ell_2,r_1)$ be these two edges. See \Cref{fig:slope-set2.2} for a schematic illustration.  Up to a renaming, we can assume that walking counter-clockwise along the outer boundary of $G$ we encounter $r_1$, $\ell_2$, $r_2$, and $\ell_1$ in this order. Let $G_1$ and $G_2$ be the two path-trees obtained by removing $e$, $e'$, and $e''$ from $G$, such that $G_i$ contains $\rho_i$, $\ell_i$, and $r_i$, for $i=1,2$. Let $T_i$ be the tree rooted at $\rho_i$ obtained by removing the edges of $G_i$ connecting its path-vertices, for $i=1,2$. Tree $T_i$ is ordered according to the embedding of $G_i$ and therefore its leftmost leaf is $\ell_i$ and its rightmost leaf is $r_i$.

We now explain how to construct a planar straight-line drawing $\Gamma$ of $G$ that uses $\Delta$ slopes. Let $\bigblacktriangle(a_1b_1c_1)$ and $\bigblacktriangle(a_2b_2c_2)$ be two equilateral triangles of the same size. By \Cref{lem:tree-in-triangle}, $T_i$, for $i=1,2$, admits a straight-line order preserving drawing $\Gamma_i$  contained in $\bigblacktriangle(a_ib_ic_i)$ with the additional properties listed in the statement of \Cref{lem:tree-in-triangle}. We rotate  $\Gamma_2$ by $\pi$ radians and translate it in such a way that the roots of $\Gamma_1$ and $\Gamma_2$ are vertically aligned (see \Cref{fig:slope-set2.c}). Notice that, since the two triangles $\bigblacktriangle(a_1b_1c_1)$ and $\bigblacktriangle(a_2b_2c_2)$ have the same size, $\ell_1$ and $r_2$ are vertically aligned and $r_1$ and $\ell_2$ are also vertically aligned. 
It follows that the edges $e=(\rho_1,\rho_2)$, $e'=(\ell_1,r_2)$, and $e''=(\ell_2,r_1)$ can be added to the drawing as vertical segments. To complete the drawing of $G$, it only remains to add the edges of the outer boundary different from $e'$ and $e''$. Since these edges only connect leaves of $T_1$ or leaves of $T_2$, and the leaves in each of such trees are horizontally aligned by \Cref{lem:tree-in-triangle}, all these edges can be drawn as horizontal segments. 

Since the two drawings use the same set of slopes $\mathcal{S}_{k}$ with $k=\max\{4,\Delta\}$ and the rotation of $\Gamma_2$ by $\pi$ radians preserves the slopes, the statement follows.\qed %

\section{Conclusions and Open Problems}\label{se:open}

In this paper we proved a quadratic upper bound on the planar slope number of nested pseudotrees. This is the first result proving the existence of graphs with treewidth $4$ whose plane slope number is polynomial in $\Delta$. In the special case of Halin graphs (which have treewidth $3$) we have an asymptotically tight $\Theta(\Delta)$ bound, which improves over the previously known $O(\Delta^5)$ bound. Our proofs are constructive and exploit the SPQ-tree, a data structure that we prove can be computed in linear time. The number of operations that we perform is also linear, however we use irrational slopes which may give rise to drawings whose area is not polynomial in the input size.

It remains open whether the same upper bounds on the slope number can be achieved if the vertices are required to lie on an integer grid of polynomial size.

Also it would be interesting to establish whether the upper bound of \cref{thm:main} is tight and whether it also applies to nested pseudoforests. Finally, 
is there a subexponential upper bound on the planar slope number of $2$-outerplanar graphs? This question is interesting even for $2$-connected graphs.


\bibliographystyle{abbrvurl}
\bibliography{main}

\end{document}